\documentclass[reqno,12pt]{article}
\usepackage{amsmath,amsfonts,amssymb,amsthm,amstext,amscd,eucal,xcolor}
\usepackage{color}
\usepackage[all]{xy}
\usepackage{cite,hyperref}
\usepackage{graphicx}
\usepackage{epsfig,epstopdf}
\usepackage[active]{srcltx}
\makeatletter \@addtoreset{equation}{section}

\makeatletter\renewcommand\section{\@startsection {section}{1}{\z@}%
                                   {-3.5ex \@plus -1ex \@minus -.2ex}%nn
                                   {2.3ex \@plus.2ex}%
                                   {\normalfont\large\bfseries}}
\renewcommand\subsection{\@startsection{subsection}{2}{\z@}%
                                     {-3.25ex\@plus -1ex \@minus -.2ex}%
                                     {1.5ex \@plus .2ex}%
                                     {\normalfont\bfseries}}

\newcommand{\be}{\begin{equation}}
\newcommand{\ee}{\end{equation}}
\newcommand{\bea}{\begin{eqnarray}}
\newcommand{\eea}{\end{eqnarray}}
\newcommand{\bse}{\begin{subequations}}
\newcommand{\ese}{\end{subequations}}
\newcommand{\beqa}{\begin{eqnarray}}
\newcommand{\eeqa}{\end{eqnarray}}
\newcommand{\beqar}{\begin{eqnarray*}}
\newcommand{\eeqar}{\end{eqnarray*}}
\newcommand{\bi}{\begin{itemize}}
\newcommand{\ei}{\end{itemize}}
\newcommand{\bn}{\begin{enumerate}}
\newcommand{\en}{\end{enumerate}}
\newcommand{\ba}{\begin{array}}
\newcommand{\ea}{\end{array}}
\newcommand{\bc}{\begin{center}}
\newcommand{\ec}{\end{center}}
\newcommand{\nnr}{\nonumber \\}

\definecolor{darkgreen}{rgb}{0,0.3,0}
\definecolor{darkblue}{rgb}{0,0,0.3}
\definecolor{darkred}{rgb}{0.7,0,0}

\newcommand{\NC}{noncommutative }
\newcommand{\email}[1]{\footnote{E-mail: \href{mailto:#1}{#1}}}
\begin{document}

\title{\textmd{\textbf{\large
{Noncommutative Maxwell-Chern-Simons theory (I): \\One-loop
dispersion relation analysis}}}}

\author{\textbf{M.~Ghasemkhani}$^{1}$\email{ghasemkhani@ipm.ir}  \textbf{and R.~Bufalo}$^{2}$
\email{rbufalo@ift.unesp.br}\\\\
 \textit{$^{1}$}
 \textit{\small{ Department of Physics, Shahid
Beheshti University,}}\\
\textit{\small{ G.C., Evin, Tehran 19839, Iran}}\\
\textit{\small{ $^{2}$ Instituto de F\'{i}sica Te\'orica,
Universidade Estadual Paulista}}\\
\textit{\small{Rua Dr. Bento Teobaldo Ferraz 271, Bloco
II Barra Funda, S\~ao Paulo, 01140-070, SP, Brazil}}\\
}
 \maketitle
\begin{abstract}
\noindent
 In this paper, we study the three-dimensional
\NC Maxwell-Chern-Simons theory. In the present analysis, a complete
account for the gauge field two-point function renormalizability is
presented and physical significant quantities are carefully
established. The respective form factor expressions from the gauge
field self-energy are computed at one-loop order. More
importantly, an analysis of the gauge field dispersion relation, in
search of possible \NC anomalies and infrared finiteness, is
performed for three special cases, with particular interest in the
highly \NC limit.
\end{abstract}
\begin{flushleft}
\end{flushleft}

%%%%%%%%%%%%%%%%%%%%%%%%%%%%%%%%%%%%%%%%%%%%%%%%%%%%%%%%%%%%%%%%%%%%%%%%%%%%%%%%%%%%%%%%%%%%%%%%%%%%%%%%%%%%%%%%%%%%%%%%%%%%%%%%%%%%%%%%%%%%%%%%%%%%%%%%%%%
\newpage
\tableofcontents

%%%%%%%%%%%%%%%%%%%%%%%%%%%%%%%%%%%%%%%%%%%%%%%%%%%%%%%%%%%%%%%%%%%%%%%%%%%%%%%%%%%%%%%%%%%%%%%%%%%%%%%%%%%%%%%%%%%%%%%%%%%%%%%%%%%%%%%%%%%%%%%%%%%%%%%%%%%
%%%%%%%%%%%%%%%%%%%%%%%%%%%%%%%%%%%%%%%%%%%%%%%%%%%%%%%%%%%%%%%%%%%%%%%%%%%%%%%%%%%%%%%%%%%%%%%%%%%%%%%%%%%%%%%%%%%%%%%%%%%%%%%%%%%%%%%%%%%%%%%%%%%%%%%%%%%
\newpage

\section{Introduction}
\label{sec:1}

In our search for a better understanding of physical
phenomena in nature, we have faced many drawbacks in explaining
well-recognized four-dimensional problems; in particular, it is of
higher importance to understand why physical nature dwells in
four dimensions. In our attempts to come close to answering these and
other questions, we have employed means that are sufficiently
intricate so that it has proven useful to wander into
lower-dimensional spacetime models \cite{Jackiw:1989nq}. Although
our initial hope was solely motivated by the wishful thought that we
could learn useful things  in a simpler setting, the wandering into
lower-dimensional models has proven to be very fertile and has
stimulated significantly the development of our knowledge in such a
way that we are now able to explain statistical systems and
condensed matter physics by means of planar physics -- in
two dimensions and three dimensions.

In recent years, we have witnessed a major advance in
the description of so many important condensed matter phenomena by
means of connections with high-energy theories
\cite{Nastase:2015wjb}. We can cite BCS superconductivity
\cite{Tinkham} and novel materials such as graphene
\cite{Vozmediano} and Weyl semimetals \cite{burkov} as some of the
most remarkable examples where a partial or fully description is
obtained by means of the use of effective proposals of gauge field
theory models \cite{Miransky:2015ava}. In particular, within the
broad class of effective models, one can find proposals in which the
violation of Lorentz symmetry is analyzed in some materials by means
of effective low-energy theories; e.g., a Lorentz-violating effective
version of QED is used to describe Weyl semimetals
\cite{Grushin:2012mt}.

 In addition to the technological advance motivated by
the development of new material and matter states, we do also live
nowadays in a thrilling era of rich high-precision experiments in
particle physics, testing long-dated gauge theories, where
structural pillars of theoretical gauge theories are scrutinized, in
particular the CPT theorem and Lorentz symmetry \cite{ref53}, whose
violation would be a sensitive signal for unconventional underlying
physics. Furthermore, if we enlarge our scope and add to our
interest the description of the nature behavior at shortest
distances \cite{ref23,ref24}, i.e. a quantum theory of gravity, or
even the so-called minimal length scale physics, one inexorably
finds that \NC geometry is one of the highly motivated and richer
frameworks \cite{ref16}, including phenomenological inspirations
\cite{ref17,Nicolini:2008aj}. In attempts to accommodate quantum
mechanics and general relativity within a common framework
\cite{doplicher}, one finds uncertainty principles that are
compatible with noncommuting coordinates, showing that the
spacetime noncommutativity naturally emerges at Plank scale.

It is well known that \NC geometry is a self-sufficient
theory, which over the past decade has found motivation
in several theoretical frameworks
\cite{connes,ho,krogh,witten,ardalan-1}, but one should highlight
its prominent role in the many phenomenological attempts to
detect sensitive deviations originated from physics at the Planck
scale \cite{ref17,ref23,ref24}. In summary, in the \NC scenario, it
is supposed that the spacetime coordinate operators do not commute
with each other and satisfy the commutation relation
\begin{equation}
[\hat{x}^{\mu},\hat{x}^{\nu}]=i\theta^{\mu\nu}, \label{eq:0.1}
\end{equation}
 in which $\theta^{\mu\nu}$ is a constant antisymmetric matrix of dimension of
length squared. To construct a \NC field theory, using the
Weyl-Moyal (symbol) correspondence \cite{alvarez}, the ordinary
product is replaced by the Moyal star product defined as
 \begin{equation}
f(x)\star
g(x)=f(x)\exp\left(\frac{i}{2}\theta^{\mu\nu}\overleftarrow{\partial_{\mu}}\overrightarrow{\partial_{\nu}}\right)
g(x).
 \end{equation}
Inserting the above star product into the Lagrangian density of the
ordinary field theory yields a highly nonlocal theory, including
higher-derivative terms which are not present in commutative theory.
Furthermore, the study of NC gauge theories have uncovered several
interesting properties. In particular, a common feature in these theories is that the
high-momentum modes (UV) affect the physics at large
distances (IR) leading to the appearance of the so-called
UV/IR mixing \cite{minwalla}, even in theories with massive particles.
Contrary to the initial expectation (which was that noncommutativity
could render UV finite field theories), this mixing complicates the
renormalization of the theory. Despite the many attempts to cure
it \cite{grosse-1,grosse-2}, with no complete success, the
problem has not yet been fully understood.

In the exact same way as we have discussed above, the
\NC three-dimensional field theory, in particular gauge theory, can
 find application in the study of planar physics in condensed
matter and statistical physics
\cite{susskind,Hellerman:2001rj,Xiao,Bastos:2012kh,Sidharth:2014xya}.
Despite the fact that some perturbative aspects of the \NC
three-dimensional field theory have been studied in the context of the
Chern-Simons theory \cite{chen,bichl,grandi,das,jabbari-0} and QED3
\cite{Blaschke:2005dv}, to the best of our knowledge, none of the
aforementioned studies were concerned with the analysis of the
anomalies that the noncommutativity can cause in the physical
content of the field, e.g. UV/IR mixing that can be present in the
physical dispersion relation of the gauge field due to radiative
corrections \cite{Matusis:2000jf}, and, therefore, modify
significantly the behavior of the quantum field in the description
of a given phenomenon \cite{Seiberg:2000gc}. In addition, this
calculation allows also an analysis regarding the infrared
finiteness of the given cases \cite{pisarski}.

The Maxwell-Chern-Simons theory consists of an
important model with the striking feature of allowing a massive
gauge field theory without any gauge symmetry breaking
\cite{jackiw}, in this case we have the so-called topologically
massive electrodynamics (it describes a helicity $\pm 1$ mode). The
presence of commutative (\NC) Chern-Simons action can also be seen
as resulting from quantum effects, arising from integrating out the
fermionic fields in commutative (\NC) massive $QED_{3}$
\cite{redlich-1,banerjee}. Moreover, we can refer to some analysis,
with a different scope than ours, in regard to the \NC
Maxwell-Chern-Simons theory \cite{dayi,Harikumar:2005ry} and its
supersymmetric extension \cite{caporaso,Ferrari}, as well as to the
higher-derivative extensions of the Chern-Simons action (in both
commutative and \NC space) \cite{deser,Kaparulin,bufalo}.

In this paper, we discuss the gauge field two-point function
renormalizability and physically significant quantities on the
one-loop order polarization tensor of the three-dimensional \NC
Maxwell-Chern-Simons theory, with particular interest in analyzing
the gauge field dispersion relation in search of possible \NC
anomalies and infrared finiteness. We begin, in Sec. \ref{sec:2}, by
reviewing the general properties of the gauge-invariant
noncommutative Maxwell-Chern-Simons theory as well as its discrete
symmetries. We determine the one-loop $1PI$ self-energy function,
and by considering a general tensor form for it, we are able to find
relations for the respective form factors. Moreover, in Sec.
\ref{sec:3}, the renormalizability for the gauge field two-point
function in this model is carefully established and afterwards
analyzed, since it can be jeopardized by the UV/IR mixing
\cite{Girotti:2000gc}. Within this context, a multiplicative
renormalization holds, and quantities of physical significance are
readily defined. In section \ref{sec:4}, we compute explicitly the
planar and nonplanar contributions for the form factor expressions,
where the commutative limit of the given outcome is investigated.
Finally, in Sec. \ref{sec:5}, we establish three particular physical
cases of interest. In particular, we examine the highly \NC limit,
where its physical dispersion relation is discussed. In Sec.
\ref{sec:6}, we summarize the results, and present our final
remarks.

%%%%%%%%%%%%%%%%%%%%%%%%%%%%%%%%%%%%%%%%%%%%%%%%%%%%%%%%%%%%%%%%%%%%%%%%%%%%%%%%%%%%%%%%%%%%%%%%%%%%%%%%%%%%%%%%%%%%%%%%%%%%%%%%%%%%%%%%%%%%%%%%%%%%%%%%%%%
%%%%%%%%%%%%%%%%%%%%%%%%%%%%%%%%%%%%%%%%%%%%%%%%%%%%%%%%%%%%%%%%%%%%%%%%%%%%%%%%%%%%%%%%%%%%%%%%%%%%%%%%%%%%%%%%%%%%%%%%%%%%%%%%%%%%%%%%%%%%%%%%%%%%%%%%%%%
\section{General remarks}
\label{sec:2}

 We start our analysis by considering the
gauge-invariant Lagrangian density of the \NC
Maxwell-Chern-Simons theory in a Minkowski spacetime
\begin{align}
\mathcal{L}=-\frac{1}{4}F_{\mu\nu}\star
F^{\mu\nu}+\frac{m}{2}\epsilon^{\mu\nu\lambda}\left(A_{\mu}\partial_{\nu}A_{\lambda}+\frac{2e}{3}A_{\mu}\star
A_{\nu}\star A_{\lambda}\right)+{\cal L}_{g.f}+{\cal L}_{gh}.
\label{eq:1.1}
\end{align}
where, the field strength tensor is
$F_{\mu\nu}=\partial_{\mu}A_{\nu}-\partial_{\nu}A_{\mu}+ie\left[A_{\mu},A_{\nu}\right]_{\star}$.
The gauge-fixing term is chosen as to the usual Lorentz condition
\[
{\cal L}_{g.f}=\frac{\xi}{2}B \star B +B \star
\left(\partial_{\mu}A^{\mu}\right), \label{eq:1.2}
\]
where $B$ is the Nakanishi-Lautrup auxiliary field and the ghost term reads
\[
{\cal L}_{gh}=\partial^{\mu}\overline{c}\star D_{\mu}^{\star}c, \label{eq:1.3}
\]
where the covariant derivative is defined such as $D_{\mu}^{\star}c=\partial_{\mu}c-ie\left[A_{\mu},c\right]_{\star}$.
The full theory \eqref{eq:1.1} is invariant under the BRST Slavnov transformations:
\begin{equation}
s A_\mu = D_{\mu}^{\star}c, \quad s c = i e~ c \star c, \quad s \overline{c} = -B, \quad s B = 0. \label{eq:1.4}
\end{equation}
The auxiliary field $B$ can be integrated out, since it plays no
part on the theory's dynamics. Now, the tree-level propagator for
the gauge field can be readily obtained, in the Landau gauge
$\xi=0$, as
\begin{align}
D_{\mu\nu}(p)=\frac{-i}{p^{2}(p^{2}-m^{2})}(p^{2}\eta_{\mu\nu}-p_{\mu}p_{\nu}+im\epsilon_{\mu\nu\lambda}p^{\lambda})
& , \label{eq:1.5}
\end{align}
where $m^{2}$ is the gauge field mass originating from the Chern-Simons term.

By completeness, in order to discuss the one-loop structure of the
polarization tensor, it is useful to review on the discrete symmetries
of parity ($\mathbf{P}$), charge conjugation ($\mathbf{C}$) and time
reversal ($\mathbf{T}$), for a three-dimensional \NC
spacetime \cite{jackiw,jabbari-1}:
%%%%%%%%%%%%%%%%%%%%%%%
\begin{itemize}
  \item (i) \emph{Parity}
\end{itemize}

    Parity transformation in $2+1$ dimensions is indeed a reflection described by $x_{1}\rightarrow -x_{1}$ and $x_{2}\rightarrow
  x_{2}$. Under parity, the gauge field transforms as
  \begin{equation}
A^{0}\rightarrow A^{0},\quad A^{1}\rightarrow -A^{1},\quad
A^{2}\rightarrow A^{2},
  \end{equation}
  which leads to a $\mathbf{P}$-invariant \NC Maxwell term
  if we consider that the parameter $\theta$ is not changed under a parity transformation.
  However, the Chern-Simons kinetic term changes sign under
  $\mathbf{P}$,
 \begin{equation}
\epsilon^{\mu\nu\lambda}A_{\mu}\partial_{\nu} A_{\lambda}\rightarrow
-\epsilon^{\mu\nu\lambda}A_{\mu}\partial_{\nu} A_{\lambda},
  \end{equation}
  whereas, for the interaction term of the Chern-Simons part, we obtain
\begin{equation}
\epsilon^{\mu\nu\lambda}A_{\mu}\star A_{\nu}\star
A_{\lambda}\rightarrow -\epsilon^{\mu\nu\lambda}A_{\mu}\star A_{\nu}
\star A_{\lambda}.
  \end{equation}
  It is thus concluded that the total \NC Chern-Simons terms are $\mathbf{P}$-odd.
  %%%%%%%%%%%%%%%%%%%%%%%%%%%%%%
\begin{itemize}
  \item (ii)\emph{ Charge conjugation}
\end{itemize}

  Under a charge conjugation transformation, the gauge field
  changes as $A_{\mu}\rightarrow -A_{\mu}$ and consequently the
  \NC Maxwell term is not $\mathbf{C}$-invariant unless we
  consider $\theta\rightarrow -\theta$, which has an intuitive
  explanation discussed in \cite{jabbari-2}. Furthermore, the Chern-Simons kinetic term transforms as
   \begin{equation}
\epsilon^{\mu\nu\lambda}A_{\mu}\partial_{\nu} A_{\lambda}\rightarrow
\epsilon^{\mu\nu\lambda}A_{\mu}\partial_{\nu} A_{\lambda}.
  \end{equation}
  To study the $\mathbf{C}$ transformation of the Chern-Simons interaction part, it is useful to rewrite it as
   \begin{equation}
\epsilon^{\mu\nu\lambda}A_{\mu}\star A_{\nu}\star A_{\lambda}=\frac{1}{2}
\epsilon^{\mu\nu\lambda}A_{\mu}\star[ A_{\nu},A_{\lambda}]_{\star},
  \end{equation}
  and therefore we have that under a charge conjugation transformation
    \begin{equation}
\epsilon^{\mu\nu\lambda}A_{\mu}\star A_{\nu}\star A_{\lambda}\rightarrow  \epsilon^{\mu\nu\lambda}A_{\mu}\star A_{\nu}\star A_{\lambda}.
  \end{equation}
  Accordingly, under the above consideration, we see that the \NC Chern-Simons term is $\mathbf{C}$-even.
  %%%%%%%%%%%%%%%%%%%%%%%
\begin{itemize}
  \item (iii) \emph{Time reversal}
\end{itemize}

  Under a time reversal transformation, the gauge field now changes as
   \begin{equation}
A^{0}\rightarrow A^{0},\quad A^{1}\rightarrow -A^{1},\quad A^{2}\rightarrow -A^{2},
  \end{equation}
  which yields a $\mathbf{T}$-invariant \NC Maxwell term, with the condition $\theta\rightarrow -\theta$.
  For the Chern-Simons free part, we obtain
     \begin{equation}
\epsilon^{\mu\nu\lambda}A_{\mu}\partial_{\nu} A_{\lambda}\rightarrow
-\epsilon^{\mu\nu\lambda}A_{\mu}\partial_{\nu} A_{\lambda},
  \end{equation}
  as well as
  \begin{equation}
\epsilon^{\mu\nu\lambda}A_{\mu}\star A_{\nu}\star
A_{\lambda}\rightarrow -\epsilon^{\mu\nu\lambda}A_{\mu}\star A_{\nu}
\star A_{\lambda}.
  \end{equation}
  Therefore, the \NC Chern-Simons term is also $\mathbf{T}$-odd.
  In view of the above arguments on discrete symmetries, we conclude that the \NC
  Maxwell action is even under $\mathbf{CP}$ and $\mathbf{PT}$, while the \NC Chern-Simons action
  is $\mathbf{CP}$-odd and $\mathbf{PT}$-even, although both of these actions are separately $\mathbf{CPT}$ invariant.
We expect that the tensor structure of the photon polarization
tensor, which is induced by quantum effects, inherits (respects) the
properties, gauge and discrete symmetries, from the classical theory.

\begin{figure}[t]
\vspace{-0.8cm}
\includegraphics[width=15cm,height=10cm]{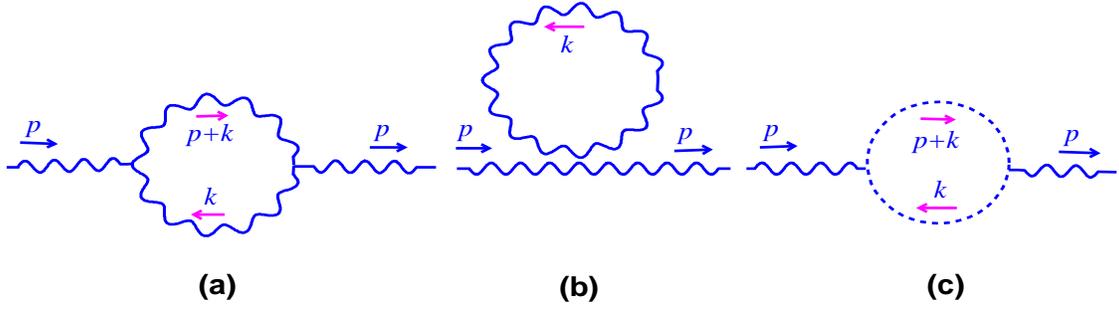}
\vspace{-4cm}
 \centering \protect\protect\protect\caption{One-loop Feynman diagrams in noncommutative
Maxwell-Chern-Simons theory: (a) gauge loop, (b) tadpole loop, (c)
ghost loop.} \label{oneloopdiagrams}
\end{figure}

The one-loop contributions for the gauge field self-energy are those from the cubic
and tadpole self-interaction and ghost loop, and these diagrams are
depicted in Fig.\ref{oneloopdiagrams}. A detailed account for each one of
them can be found at Appendix \ref{appA}. These contributions can be
conveniently written in the following form \eqref{eq:A.4},
\begin{align}
\Pi_{\mu\nu}\left(p\right)=e^{2}\int\frac{d^{3}k}{(2\pi)^{3}}~\sin^{2}\left(\frac{p\wedge k}{2}\right)~
\frac{{\cal N}_{\mu\nu}^{\mathrm{g}}+2{\cal N}_{\mu\nu}^{\mathrm{gh}}+2{\cal N}_{\mu\nu}^{\mathrm{t}}}{k^{2}(k^{2}-m^{2})
(p+k)^{2}((p+k)^{2}-m^{2})}, \label{eq:1.6}
\end{align}
where we have used the notation $p\wedge k = p_{\mu}\theta^{\mu \nu}k_{\nu}$. Also, the tensor quantities at the numerator
are defined, respectively, by Eq.\eqref{eq:A.3},
\begin{align}
{\cal N}_{\mu\nu}^{\mathrm{g}} & =\left(im\epsilon_{\mu\alpha\beta}+(p+2k)_{\mu}\eta_{\alpha\beta}+(p-k)_{\beta}\eta_{\mu\alpha}-(2p+k)_{\alpha}\eta_{\mu\beta}\right)\nonumber \\
 & \times\left(im\epsilon_{\nu\rho\sigma}-(p+2k)_{\nu}\eta_{\rho\sigma}+(k-p)_{\sigma}\eta_{\rho\nu}+(2p+k)_{\rho}\eta_{\nu\sigma}\right) \\
 & \times\left(k^{2}\eta^{\alpha\rho}-k^{\alpha}k^{\rho}+im\epsilon^{\alpha\rho\lambda}k_{\lambda}\right)\left((p+k)^{2}\eta^{\beta\sigma}-(p+k)^{\beta}(p+k)^{\sigma}-im\epsilon^{\beta\sigma\xi}(p+k)_{\xi}\right),\nonumber
\end{align}
and Eqs.\eqref{eq:A.5} and \eqref{eq:A.6}
\begin{align}
{\cal N}_{\mu\nu}^{\mathrm{gh}} & =m^{4}\left(k_{\mu}k_{\nu}+k_{\mu}p_{\nu}\right)-m^{2}\left(2k^{2}+p^{2}+2p.k\right)\left(k_{\mu}k_{\nu}+k_{\mu}p_{\nu}\right)\nonumber \\
 & +k^{2}\left(k^{2}+2p.k+p^{2}\right)\left(k_{\mu}k_{\nu}+k_{\mu}p_{\nu}\right),\\
{\cal N}_{\mu\nu}^{\mathrm{t}} & =-m^{2}\left(k^{2}+2p.k+p^{2}\right)\left(k^{2}\eta_{\mu\nu}+k_{\mu}k_{\nu}\right)\nonumber \\
 & +\left(k^{4}+2k^{2}p^{2}+p^{4}+4k^{2}(p.k)+4p^{2}(p.k)+4(p.k)^{2}\right)\left(k^{2}\eta_{\mu\nu}+k_{\mu}k_{\nu}\right).
\end{align}

As a check for Eq.\eqref{eq:1.6}, we see that, apart from the
trigonometric factor $\sin^{2}\left(\frac{p\wedge k}{2}\right)$, the
remaining of the expression is exactly the same as the one appearing
in \cite{pisarski}, where a detailed one-loop analysis of the
Yang-Mills-Chern-Simons theory is presented.
 The most general tensor structure of the photon
self-energy in a \NC three-dimensional spacetime is given as
\begin{align}
\Pi^{\mu\nu}= & \bigg(\eta^{\mu\nu}-\frac{p^{\mu}p^{\nu}}{p^{2}}\bigg)\Pi^{\star}_{\mathrm{e}}+
\frac{\tilde{p}^{\mu}\tilde{p}^{\nu}}{\tilde{p}^{2}}~\widetilde{\Pi}^{\star}_{\mathrm{e}}
+i\Pi_{\mathrm{o}}^{\mathrm{\tiny{A}}}\epsilon^{\mu\nu\lambda}p_{\lambda}
+\Pi_{\mathrm{o}}^{\mathrm{\tiny{S}}}\bigg(\tilde{p}^{\mu}u^{\nu}+\tilde{p}^{\nu}u^{\mu}\bigg), \label{eq:1.7}
\end{align}
where we had chosen
$u_{\mu}=\epsilon_{\mu\alpha\beta}p^{\alpha}\tilde{p}^{\beta}$, with
$p^{\mu}u_{\mu}=\tilde{p}^{\mu}u_{\mu}=0$ as our orthonormal basis.
We notice, however, that $\Pi^{\mu\nu}$ in this basis has nine terms
that reduce to four terms due to the Ward identity \footnote{Since the
tensor $\Pi^{\mu\nu}$ is not totally symmetric, the Ward identity
must hold for both $p_{\mu}\Pi^{\mu\nu}=0$ and
$p_{\nu}\Pi^{\mu\nu}=0$.}(for further details, see Appendix
\ref{appB}). It is notable that the tensor structure of the first
and the third term in \eqref{eq:1.7} is analogous to that of the
tree-level counterparts in commutative Maxwell-Chern-Simons model,
which is a free theory. Indeed, the first term is even under
$\mathbf{CP}$ and $\mathbf{PT}$, while the third term is
$\mathbf{CP}$-odd and $\mathbf{PT}$-even, they are similar to the
commutative Maxwell and Chern-Simons actions, respectively.

On the other hand, the second and the fourth terms have no tree-level
counterparts in commutative Maxwell-Chern-Simons model, arising
fully from quantum effects. Furthermore, the second term in
\eqref{eq:1.7}, similar to the first term, is even under
$\mathbf{CP}$ and $\mathbf{PT}$, while   the fourth term, similar to
the third term, is $\mathbf{CP}$-odd and $\mathbf{PT}$-even.
Consequently, the behavior of the quantum effect terms is the same
as that of the tree-level terms, as we expected.
 We observe hence that the specific decomposition appearing
in the tensor structure of the photon self-energy in \eqref{eq:1.7} is
physically justified, using the aforementioned discussion on
discrete symmetries.

Besides, we see that the one-loop photon self-energy \eqref{eq:1.6}
is invariant under $\theta\rightarrow -\theta$, which is in
agreement with its tensor structure described by \eqref{eq:1.7}.
Consequently, all of the form factor coefficients
$\Pi^{\star}_{\mathrm{e}}$, $\widetilde{\Pi}^{\star}_{\mathrm{e}}$,
$\Pi_{\mathrm{o}}^{\mathrm{\tiny{A}}}$ and
$\Pi_{\mathrm{o}}^{\mathrm{\tiny{S}}}$ in \eqref{eq:1.7} are
expected to be even in $\theta$, at least at one-loop level, as we
see in the following identities,
Eqs.\eqref{eq:B.6a}--\eqref{eq:B.6d},
\begin{align}
\Pi^{\star}_{\mathrm{e}}= & \eta_{\mu\nu}\Pi^{\mu\nu}-\frac{\tilde{p}_{\mu}\tilde{p}_{\nu}}{\tilde{p}^{2}}\Pi^{\mu\nu},\label{eq:1.8a}\\
\widetilde{\Pi}^{\star}_{\mathrm{e}}= & -\eta_{\mu\nu}\Pi^{\mu\nu}+2\frac{\tilde{p}_{\mu}\tilde{p}_{\nu}}{\tilde{p}^{2}}\Pi^{\mu\nu},\label{eq:1.8b}\\
\Pi_{\mathrm{o}}^{\mathrm{\tiny{A}}}= & \frac{i}{2p^{2}}\epsilon_{\mu\nu\alpha}p^{\alpha}\Pi^{\mu\nu},\label{eq:1.8c}\\
\Pi_{\mathrm{o}}^{\mathrm{\tiny{S}}}= & -\frac{1}{2\tilde{p}^{4}p^{2}}\left(u_{\mu}\tilde{p}_{\nu}+u_{\nu}\tilde{p}_{\mu}\right)\Pi^{\mu\nu}.
\label{eq:1.8d}
\end{align}

With these quantities we have introduced all necessary information on regard to our analysis.
Next, we shall proceed and write explicitly the one-loop expressions for the form factors of the gauge field self-energy.
%%%%%%%%%%%%%%%%%%%%%%%%%%%%%%%%%%%%%%%%%%%%%%%%%%%%%%%%%%%%%%%%%%%%%%%%%%%%%%%%%%%%%%%%%%%%%%%%%%%%%%%%%%%%%%%%%%%%%%%%%%%%%%%%%%%%%%%%%%%%%%%%%%%%%%%%%%%
%%%%%%%%%%%%%%%%%%%%%%%%%%%%%%%%%%%%%%%%%%%%%%%%%%%%%%%%%%%%%%%%%%%%%%%%%%%%%%%%%%%%%%%%%%%%%%%%%%%%%%%%%%%%%%%%%%%%%%%%%%%%%%%%%%%%%%%%%%%%%%%%%%%%%%%%%%%
\subsection{Form factors}
\label{sec:2.1}

In order to evaluate the above relations, Eqs.\eqref{eq:1.8a}--\eqref{eq:1.8d},
we shall take the respective tensor contraction with the one-loop expression \eqref{eq:1.6}.
First, the relation \eqref{eq:1.8a} yields the following result,
\begin{align}
\Pi^{\star}_{\mathrm{e}}\left(p^{2}\right)&=i  e^{2}\int\frac{d^{3}k}{(2\pi)^{3}}~\sin^{2}\left(\frac{p\wedge k}{2}\right)~\frac{1}{\left(p+k\right)^{2}\left(\left(p+k\right)^{2}-m^{2}\right)}\nonumber \\
&\times \biggl\{6(p.k)+4\frac{(k.\tilde{p})^{2}}{\tilde{p}^{2}}-16m^{2}+\frac{1}{\left(k^{2}-m^{2}\right)}\biggl(16(p.k)^{2}+10p^{2}(p.k)+p^{4}\nonumber \\
 & +\frac{(k.\tilde{p})^{2}}{\tilde{p}^{2}}\left(4(p.k)-2p^{2}+16m^{2}\right)-26m^{2}(p.k)-12m^{2}p^{2}-16m^{4}\biggr)\nonumber \\
 & +\frac{1}{k^{2}\left(k^{2}-m^{2}\right)}\biggl(\frac{(k.\tilde{p})^{2}}{\tilde{p}^{2}}\left(m^{2}\left(8p^{2}+14(p.k)\right)-4(p.k)^{2}-2p^{4}-8p^{2}(p.k)\right)\nonumber \\
 &-4m^{2}p^{2}(p.k)-8m^{2}(p.k)^{2}+3p^{2}(p.k)^{2}+4(p.k)^{3}\biggr)\biggr\},\label{eq:1.10}
\end{align}
Besides, from the relation \eqref{eq:1.8b}, we obtain
\begin{align}
\widetilde{\Pi}^{\star}_{\mathrm{e}}\left(p^{2}\right) & =ie^{2}\int\frac{d^{3}k}{(2\pi)^{3}}~\sin^{2}\left(\frac{p\wedge k}{2}\right)~\frac{1}{\left(p+k\right)^{2}\left(\left(p+k\right)^{2}-m^{2}\right)}\nonumber \\
 & \times\biggl\{2k^{2}+2(p.k)-4p^{2}+16m^{2}-8\frac{(k.\tilde{p})^{2}}{\tilde{p}^{2}}
 +\frac{1}{\left(k^{2}-m^{2}\right)}\biggl(m^{2}\left(30(p.k)+7p^{2}+14m^{2}\right)\nonumber \\
 &+2m^{4} -3p^{4}-10p^{2}(p.k)+\frac{(k.\tilde{p})^{2}}{\tilde{p}^{2}}\left(4p^{2}-8(p.k)-32m^{2}\right)\biggr)\nonumber \\
 & +\frac{1}{k^{2}\left(k^{2}-m^{2}\right)}\biggl[\frac{(k.\tilde{p})^{2}}{\tilde{p}^{2}}\bigg(8(p.k)^{2}+4p^{4}+16p^{2}(p.k)-2m^{2}\bigg[8p^{2}+14(p.k)\bigg]\bigg)\nonumber \\
 &+\left(p.k\right)\left(m^{2}\bigg(4p^{2}+7\left(p.k\right)\bigg)+p^{2}\left(p.k\right)+4\left(p.k\right)^{2}\right)\biggr]\biggr\}.\label{eq:1.11}
\end{align}
Moreover, from the relation \eqref{eq:1.8c}, we find the odd form factor $\Pi_{\mathrm{o}}^{\mathrm{\tiny{A}}}$ as
\begin{align}
\Pi_{\mathrm{o}}^{\mathrm{\tiny{A}}}\left(p^{2}\right) & =\frac{2mie^{2}}{p^{2}}\int\frac{d^{3}k}{(2\pi)^{3}}~\sin^{2}\left(\frac{p\wedge k}{2}\right)~\frac{1}{\left(p+k\right)^{2}\left(\left(p+k\right)^{2}-m^{2}\right)}\label{eq:1.12} \\
 & \times\biggl\{5p^{2}+\frac{\left(\left(5\left(p.k\right)+4p^{2}+7m^{2}\right)p^{2}-5\left(p.k\right)^{2}\right)}{\left(k^{2}-m^{2}\right)}-\frac{\left(p.k\right)^{2}\left( 2m^{2}+5\left(p.k\right)+4p^{2}\right) }{k^{2}\left(k^{2}-m^{2}\right)}\biggr\}.\nonumber
\end{align}
Finally, the form factor $\Pi_{\mathrm{o}}^{\mathrm{\tiny{S}}}$ follows \eqref{eq:1.8d} as
\begin{align}
\Pi_{\mathrm{o}}^{\mathrm{\tiny{S}}}\left(p^{2}\right) & =\frac{i}{\tilde{p}^{4}p^{2}}e^{2}\int\frac{d^{3}k}{(2\pi)^{3}}~\sin^{2}\left(\frac{p\wedge k}{2}\right)~\frac{\left(u.k\right)\left(\tilde{p}.k\right)}{\left(p+k\right)^{2}\left(\left(p+k\right)^{2}-m^{2}\right)}\nonumber \\
 & \times\biggl\{ 4+\frac{1}{\left(k^{2}-m^{2}\right)}\left(16m^{2}-2p^{2}+4(p.k)\right)\nonumber \\
 &+\frac{1}{k^{2}\left(k^{2}-m^{2}\right)}\left(14m^{2}(p.k)+8m^{2}p^{2}-4(p.k)^{2}-2p^{4}-8p^{2}(p.k)\right)\biggr\} .\label{eq:1.13}
\end{align}

In order to evaluate the above momentum integration, we can make use of the standard
rules for Feynman integrals. We shall next continue with our formal
development by providing now a detailed account for the renormalizability of the photon two-point function,
which analysis will allow us to define properly the one-loop dispersion relation.

%%%%%%%%%%%%%%%%%%%%%%%%%%%%%%%%%%%%%%%%%%%%%%%%%%%%%%%%%%%%%%%%%%%%%%%%%%%%%%%%%%%%%%%%%%%%%%%%%%%%%%%%%%%%%%%%%%%%%%%%%%%%%%%%%%%%%%%%%%%%%%%%%%%%%%%%%%%
%%%%%%%%%%%%%%%%%%%%%%%%%%%%%%%%%%%%%%%%%%%%%%%%%%%%%%%%%%%%%%%%%%%%%%%%%%%%%%%%%%%%%%%%%%%%%%%%%%%%%%%%%%%%%%%%%%%%%%%%%%%%%%%%%%%%%%%%%%%%%%%%%%%%%%%%%%%
\section{Renormalized gauge propagator and mass}
\label{sec:3}

We shall now formally establish the gauge field two-point function
renormalizability. In particular, we want to determine the
renormalized gauge propagator and mass, which allow us to define the
physical pole and, therefore, the dispersion relation of the gauge
field. We start by writing the complete propagator expression
\eqref{eq:B.5} \footnote{For a complete account of this discussion
see Appendix \ref{appB}. Moreover, we take
$\Pi_{\mathrm{o}}^{\mathrm{\tiny{S}}}=0$ in Eq.\eqref{eq:B.5}. This
is explicitly shown in Eqs.\eqref{eq:3.14} and \eqref{eq:3.15},
which is expected to be true to all orders.}
\begin{equation}
i\mathcal{D}_{\mu\nu}=\frac{p^{2}-\Pi^{\star}_{\mathrm{e}}-\widetilde{\Pi}^{\star}_{\mathrm{e}}}{\mathcal{R}}\left(\eta_{\mu\nu}-\frac{p_{\mu}p_{\nu}}{p^{2}}-\frac{\tilde{p}_{\mu}\tilde{p}_{\nu}}{\tilde{p}^{2}}\right)+\frac{p^{2}-\Pi^{\star}_{\mathrm{e}}}{\mathcal{R}}\frac{\tilde{p}_{\mu}\tilde{p}_{\nu}}{\tilde{p}^{2}}+\frac{m+\Pi_{\mathrm{o}}^{\mathrm{\tiny{A}}}}{\mathcal{R}}i\varepsilon_{\mu\nu\lambda}p^{\lambda}+\frac{\xi}{p^{4}}p_{\mu}p_{\nu}.\label{eq:2.1}
\end{equation}
where the quantity $\mathcal{R}$ at the denominator is given by
\[
\mathcal{R}=(p^{2}-\Pi^{\star}_{\mathrm{e}})(p^{2}-\Pi^{\star}_{\mathrm{e}}-\widetilde{\Pi}^{\star}_{\mathrm{e}})+p^{2}\bigg[(\tilde{p}^{2}\Pi_{\mathrm{o}}^{\mathrm{\tiny{S}}})^{2}-(m+\Pi_{\mathrm{o}}^{\mathrm{\tiny{A}}})^{2}\bigg].
\]

It proves to be convenient for our development to make a few replacements,
\begin{equation}
\left\{
\Pi^{\star}_{\mathrm{e}},\widetilde{\Pi}^{\star}_{\mathrm{e}}\right\}
\rightarrow p^{2}\left\{
\Pi_{\mathrm{e}},\widetilde{\Pi}_{\mathrm{e}}\right\} ,
\end{equation}
where $\Pi_{\mathrm{e}}$ and
$\widetilde{\Pi}_{\mathrm{e}}$ are dimensionless form factors.
 With this new definition, and also introducing the
notation
$\Pi_{\mathrm{e}}^{'}=\Pi_{\mathrm{e}}+\widetilde{\Pi}_{\mathrm{e}}$,
the exact propagator \eqref{eq:2.1} is conveniently rewritten as
\begin{align}
i\mathcal{D}_{\mu\nu} & =\frac{1}{\left(1-\Pi_{\mathrm{e}}\right)\bigg[p^{2}-\frac{(m+\Pi_{\mathrm{o}}^{\mathrm{\tiny{A}}})^{2}}{(1-\Pi_{\mathrm{e}})(1-\Pi_{\mathrm{e}}^{'})}\bigg]}\left(\eta_{\mu\nu}-\frac{p_{\mu}p_{\nu}}{p^{2}}-\frac{\tilde{p}_{\mu}\tilde{p}_{\nu}}{\tilde{p}^{2}}\right)
 +\frac{1}{\left(1-\Pi_{\mathrm{e}}^{'}\right)\bigg[p^{2}-\frac{(m+\Pi_{\mathrm{o}}^{\mathrm{\tiny{A}}})^{2}}{(1-\Pi_{\mathrm{e}})(1-\Pi_{\mathrm{e}}^{'})}\bigg]}\frac{\tilde{p}_{\mu}\tilde{p}_{\nu}}{\tilde{p}^{2}}\nonumber \\
 &+\frac{\left(m+\Pi_{\mathrm{o}}^{\mathrm{\tiny{A}}}\right)}{\left(1-\Pi_{\mathrm{e}}\right)\left(1-\Pi_{\mathrm{e}}^{'}\right)\bigg[p^{2}-\frac{(m+\Pi_{\mathrm{o}}^{\mathrm{\tiny{A}}})^{2}}{(1-\Pi_{\mathrm{e}})(1-\Pi_{\mathrm{e}}^{'})}\bigg]}~\frac{i\varepsilon_{\mu\nu\lambda}p^{\lambda}}{p^{2}}+\frac{\xi}{p^{4}}p_{\mu}p_{\nu}.\label{eq:2.2}
\end{align}
From this expression, we can readily identify the respective renormalization functions and renormalized mass.
Thus, we introduce the wave function and mass renormalization constants as the following,
\begin{equation}
{\cal {Z}}=1-\Pi_{\mathrm{e}},\qquad\tilde{\mathcal{Z}}=1-\Pi_{\mathrm{e}}^{'},\qquad{\cal {Z}}_{\mathrm{m}}=1+m^{-1}\Pi_{\mathrm{o}}^{\mathrm{\tiny{A}}},\label{eq:2.3}
\end{equation}
and we also define the renormalized mass as
\begin{equation}
m_{\mathrm{ren}}^{2}=\frac{(m+\Pi_{\mathrm{o}}^{\mathrm{\tiny{A}}})^{2}}{(1-\Pi_{\mathrm{e}})
(1-\Pi_{\mathrm{e}}^{'})}=\frac{{\cal {Z}}^2_{\mathrm{m}}}{{\cal {Z}}\tilde{\mathcal{Z}}}m^{2}.\label{eq:2.4}
\end{equation}
The most significant consequence arising from the multiplicative
property of \eqref{eq:2.4} is that the gauge symmetry is exactly
preserved at classical (tree) and quantum (loop) levels. Mainly
because $m=0$ corresponds to the \NC Maxwell theory which
is gauge invariant at any order, without any mass generation. Hence,
by taking into account the above definitions, Eqs.\eqref{eq:2.3} and
\eqref{eq:2.4}, we rewrite the propagator \eqref{eq:2.2} in the form
\begin{align}
i\mathcal{D}_{\mu\nu} & =\frac{1}{{\cal {Z}}\left[p^{2}-m_{\mathrm{ren}}^{2}\right]}\left(\eta_{\mu\nu}-\frac{p_{\mu}p_{\nu}}{p^{2}}-\frac{\tilde{p}_{\mu}\tilde{p}_{\nu}}{\tilde{p}^{2}}\right)\nonumber \\
&+\frac{1}{\tilde{\mathcal{Z}}\left[p^{2}-m_{\mathrm{ren}}^{2}\right]}\frac{\tilde{p}_{\mu}\tilde{p}_{\nu}}{\tilde{p}^{2}}+\frac{m_{\mathrm{ren}}}{\sqrt{{\cal {Z}}\tilde{\mathcal{Z}}}\left[p^{2}-m_{\mathrm{ren}}^{2}\right]}\frac{i\varepsilon_{\mu\nu\lambda}p^{\lambda}}{p^{2}}+\frac{\xi}{p^{4}}p_{\mu}p_{\nu}. \label{eq:2.5}
\end{align}
It should be emphasized that in view of Eq.\eqref{eq:2.5} the
multiplicative renormalization holds for the gauge-invariant \NC
Maxwell-Chern-Simons theory, which is a remarkable result.
 Furthermore, we should stress the fact that the physical massive
pole $p^{2}=m_{\mathrm{ren}}^{2}$ is found to be naturally present
at all physical terms of the complete propagator.

Since we are working within a perturbation theory approach, we can express the physically significant
renormalized mass $m_{\mathrm{ren}}$ \eqref{eq:2.4} at the lowest-order in terms of the form factors as
\begin{equation}
m_{\mathrm{ren}}=\frac{(m+\Pi_{\mathrm{o}}^{\mathrm{\tiny{A}}})}{\sqrt{(1-\Pi_{\mathrm{e}})(1-\Pi_{\mathrm{e}}^{'})}}\simeq  m\left(1+\frac{1}{m}
\Pi_{\mathrm{o}}^{\mathrm{\tiny{A}}}+ \Pi_{\mathrm{e}}+\frac{1}{2}\widetilde{\Pi}_{\mathrm{e}}+\mathcal{O}(\alpha^{2}) \right).  \label{eq:2.6}
\end{equation}
whereas the dispersion relation $p^{2}=m_{\mathrm{ren}}^{2}$, the renormalized mass at the lowest order is given by
\eqref{eq:2.6} and, hence, written in a convenient form:
\begin{equation}
\omega^{2}=|\vec{p}|^{2}+m^{2}\left(1+\frac{2}{m}\Pi_{\mathrm{o}}^{\mathrm{\tiny{A}}}
+2\Pi_{\mathrm{e}}+\widetilde{\Pi}_{\mathrm{e}}+\mathcal{O}(\alpha^{2})\right).\label{eq:2.8}
\end{equation}

With this renormalizability discussion we conclude our formal
development for NC Maxwell-Chern-Simons theory. We will proceed now
to compute explicitly the one-loop self-energy function, in
particular its form factors. Afterwards, we shall particularize the
results by considering some physical relevant limits where
analytical expressions for the dispersion relation are found.
In addition to these discussions on the dispersion
relation, we will scrutinize it for an analysis of the UV/IR mixing
in order to verify whether or not it jeopardizes the theory's
renormalizability \cite{Girotti:2000gc}.

%%%%%%%%%%%%%%%%%%%%%%%%%%%%%%%%%%%%%%%%%%%%%%%%%%%%%%%%%%%%%%%%%%%%%%%%%%%%%%%%%%%%%%%%%%%%%%%%%%%%%%%%%%%%%%%%%%%%%%%%%%%%%%%%%%%%%%%%%%%%%%%%%%%%%%%%%%%
%%%%%%%%%%%%%%%%%%%%%%%%%%%%%%%%%%%%%%%%%%%%%%%%%%%%%%%%%%%%%%%%%%%%%%%%%%%%%%%%%%%%%%%%%%%%%%%%%%%%%%%%%%%%%%%%%%%%%%%%%%%%%%%%%%%%%%%%%%%%%%%%%%%%%%%%%%%
\section{One-loop radiative correction}
\label{sec:4}

In order to compute the momentum integration on the form factors we shall make use of the standard Feynman parametrization
and dimensional regularization method.
Some relevant results for the nonplanar integration can be found at Appendix \ref{appC}, also we present the complete
expression of some lengthy expression of the form factors in Appendix \ref{appD}.

%%%%%%%%%%%%%%%%%%%%%%%%%%%%%%%%%%%%%%%%%%%%%%%%%%%%%%%%%%%%%%%%%%%%%%%%%%%%%%%%%%%%%%%%%%%%%%%%%%%%%%%%%%%%%%%%%%%%%%%%%%%%%%%%%%%%%%%%%%%%%%%%%%%%%%%%%%%
%%%%%%%%%%%%%%%%%%%%%%%%%%%%%%%%%%%%%%%%%%%%%%%%%%%%%%%%%%%%%%%%%%%%%%%%%%%%%%%%%%%%%%%%%%%%%%%%%%%%%%%%%%%%%%%%%%%%%%%%%%%%%%%%%%%%%%%%%%%%%%%%%%%%%%%%%%%
\subsection{Transverse part $\Pi_{\mathrm{e}}$}
\label{sec:4.1}

In this first study we will perform the calculation by reviewing some relevant detail.
We start by making use of the Feynman parametrization to write the denominator of Eq.\eqref{eq:1.10} in the form
\begin{align}
\Pi_{\mathrm{e}}\left(p^{2}\right) & =\frac{ie^{2}}{p^2}\int\frac{d^{3}k}{(2\pi)^{3}}~\sin^{2}\left(\frac{p\wedge k}{2}\right)\biggl\{\Gamma\left(2\right)\int d\Phi\frac{1}{\left[\left(p+k\right)^{2}-\Delta_{1}^{2}\right]^{2}}\left(6(p.k)+4\frac{(k.\tilde{p})^{2}}{\tilde{p}^{2}}-16m^{2}\right)\nonumber \\
 & +\Gamma\left(3\right)\int d\varUpsilon\frac{1}{\left[\left(k+\left(y+z\right)p\right)^{2}-\Delta_{2}^{2}\right]^{3}}\biggl(\frac{(k.\tilde{p})^{2}}{\tilde{p}^{2}}\left(4(p.k)-2p^{2}+16m^{2}\right)\nonumber \\
 &+16(p.k)^{2}+10p^{2}(p.k)+p^{4}-26m^{2}(p.k)-12m^{2}p^{2}-16m^{4}\biggr)\nonumber \\
 & +\Gamma\left(4\right)\int d\varXi\frac{1}{\left[\left(k+\left(z+w\right)p\right)^{2}-\Delta_{3}^{2}\right]^{4}}\biggl(4(p.k)^{3}
 -4m^{2}p^{2}(p.k)-8m^{2}(p.k)^{2}+3p^{2}(p.k)^{2}\nonumber \\
 &+\frac{(k.\tilde{p})^{2}}{\tilde{p}^{2}}\left(\left(8p^{2}+14(p.k)\right)m^{2}-4(p.k)^{2}-2p^{4}-8p^{2}(p.k)\right)\biggr)\biggr\}, \label{eq:3.1}
\end{align}
where we have introduced the following notation for the integration
measures,
\begin{gather}
\int d\Phi=\int dxdy\delta\left(x+y-1\right),\quad\int d\varUpsilon=\int dxdydz\delta\left(x+y+z-1\right),\nonumber \\
\int d\varXi=\int dxdydzdw\delta\left(x+y+z+w-1\right),
\end{gather}
and defined the following quantities as well:
\begin{align}
\Delta_{1}^{2} & =ym^{2}, \label{eq:3.2a}\\
\Delta_{2}^{2} & =\left(x+z\right)m^{2}-\left(y+z\right)\left(1-y-z\right)p^{2},\label{eq:3.2b}\\
\Delta_{3}^{2} & =\left(y+w\right)m^{2}-\left(z+w\right)\left(1-z-w\right)p^{2}.\label{eq:3.2c}
\end{align}
Next, by making a suitable change of variables on the momentum integration on the terms of \eqref{eq:3.1}, we find the expression
\begin{align}
\Pi_{\mathrm{e}}\left(p^{2}\right) & =\frac{i\mu^{2\left(3-\omega\right)}e^{2}}{p^2}\biggl\{\Gamma\left(2\right) \int d\Phi\left(\left(-6p^{2}-16m^{2}\right) ~ \Omega_{2} \left(\Delta_{1}\right)+4\frac{\tilde{p}_{\mu}\tilde{p}_{\nu}}{\tilde{p}^{2}} ~ \Omega_{2}^{\mu \nu  } \left(\Delta_{1}\right)\right)\label{eq:3.3}\\
 & +\Gamma\left(3\right)\int d\varUpsilon\biggl(\left[16\left(y+z\right)^{2}p^{4}-10\left(y+z\right)p^{4}+\left(26\left(y+z\right)-12\right)m^{2}p^{2}+p^{4}-16m^{4}\right] ~ \Omega_{3} \left(\Delta_{2}\right)\nonumber \\
 & +\left[16p_{\mu}p_{\nu}+\frac{\tilde{p}_{\mu}\tilde{p}_{\nu}}{\tilde{p}^{2}}\left(-\left(4\left(y+z\right)+2\right)p^{2}+16m^{2}\right)\right]
 ~ \Omega_{3}^{\mu \nu  } \left(\Delta_{2}\right)\biggr)\nonumber \\
 & +\Gamma\left(4\right)\int d\varXi\biggl(p^{4}\left[4\left(z+w\right)\left(1-2\left(z+w\right)\right)m^{2}+\left(z+w\right)^{2}\left(3-4\left(z+w\right)\right)p^{2}\biggr)\right]  ~ \Omega_{4} \left(\Delta_{3}\right) \nonumber \\
 & +\biggl[\left(m^{2}\left(8-14\left(z+w\right)\right)p^{2}-4\left(z+w\right)^{2}p^{4}-2p^{4}+8\left(z+w\right)p^{4}\right)\frac{\tilde{p}_{\mu}\tilde{p}_{\nu}}{\tilde{p}^{2}}\nonumber \\
 &+\left(-8m^{2}+3p^{2}-12\left(z+w\right)p^{2}\right) p_{\mu}p_{\nu} \biggr]  ~ \Omega_{4}^{\mu \nu  } \left(\Delta_{3}\right) -\frac{4}{\tilde{p}^{2}}p_{\mu}p_{\nu}\tilde{p}_{\lambda}\tilde{p}_{\beta} ~ \Omega_{4}^{\mu \nu \lambda \beta} \left(\Delta_{3}\right) \biggr)\biggr\},\nonumber
\end{align}
where, by convenience, we have introduced the following notation for the momentum integration:
\begin{align}
\left\{\Omega_{a}, \Omega_{a}^{\mu \nu}, \Omega_{a}^{\mu \nu \lambda \beta} \right\} \left(\Delta_{i}\right)=\int\frac{d^{\omega}Q}{\left(2\pi\right)^{\omega}}\sin^{2}\left(\frac{p\wedge Q}{2}\right)\frac{ \left\{{\bf 1},Q^{\mu}Q^{\nu},Q^{\mu}Q^{\nu}Q^{\lambda}Q^{\beta}\right\}}{\left[Q^{2}-\Delta_{i}^{2}\right]^{a}} \label{eq:3.4}
\end{align}
The integration from \eqref{eq:3.4} can be readily calculated (see
Appendix \ref{appC}). In particular, we can separate the planar and
nonplanar contributions by using the trigonometric relation
$2\sin^{2}\left(\frac{p\wedge Q}{2}\right)=1-\cos\left(p\wedge
Q\right)$. The expressions for the planar and
nonplanar contributions are explicitly given by Eqs.\eqref{eq:3.5}
and \eqref{eq:3.6}, respectively.

A remark is in place for the expressions
Eqs.\eqref{eq:3.5} and \eqref{eq:3.6} (as well for the next ones).
Since the remaining integrals on the Feynman parameters of these
expressions are rather difficult to compute exactly (and no
substantial information would be obtained), we shall leave it only
indicated and evaluate them in some particular cases, which imply
some simplification on the integrand, and we can hence discuss some
interesting physical implications. We will also do this for the
remaining contributions.

In particular, one can realize that, at the commutative limit, i.e. $\theta \rightarrow 0$, the planar and nonplanar contributions, Eqs.\eqref{eq:3.5} and \eqref{eq:3.6}, result as expected into
\begin{align}
\left(\Pi_{\mathrm{e}}\right)_{\mathrm{p}}\left(p^{2}\right)+\lim_{\theta\rightarrow~0}
\left(\Pi_{\mathrm{e}}\right)_{\mathrm{n-p}}\left(p^{2}\right)=0.\label{eq:3.c1}
\end{align}
\noindent
Furthermore, this vanishing result is in agreement with the
tree-level (propagator) structure of the commutative
Maxwell-Chern-Simons which is a free theory.

%%%%%%%%%%%%%%%%%%%%%%%%%%%%%%%%%%%%%%%%%%%%%%%%%%%%%%%%%%%%%%%%%%%%%%%%%%%%%%%%%%%%%%%%%%%%%%%%%%%%%%%%%%%%%%%%%%%%%%%%%%%%%%%%%%%%%%%%
%%%%%%%%%%%%%%%%%%%%%%%%%%%%%%%%%%%%%%%%%%%%%%%%%%%%%%%%%%%%%%%%%%%%%%%%%%%%%%%%%%%%%%%%%%%%%%%%%%%%%%%%%%%%%%%%%%%%%%%%%%%%%%%%%%%%%%%%
\subsection{NC transverse part $\widetilde{\Pi}_{\mathrm{e}}$}
\label{sec:4.2}

Next, in order to evaluate the integration of the contribution \eqref{eq:1.11}, we follow the aforementioned steps and
 write it conveniently in terms of the quantities $ \Omega_{a}^{\mu \nu ...} \left(\Delta_{i}\right)$
\begin{align}
\widetilde{\Pi}_{\mathrm{e}}\left(p^{2}\right) & =\frac{i\mu^{2\left(3-\omega\right)}e^{2}}{p^2}\biggl\{\Gamma\left(2\right)\int d\Phi\left[\left(-4p^{2}+16m^{2}\right) ~ \Omega_{2}  \left(\Delta_{1}\right)+\left(2\eta_{\mu\nu}-8\frac{\tilde{p}_{\mu}\tilde{p}_{\nu}}{\tilde{p}^{2}}\right)~ \Omega_{2}^{\mu \nu } \left(\Delta_{1}\right)\right]\nonumber \\
 & +\Gamma\left(3\right)\int d\varUpsilon\biggl(\left(m^{2}\left(-30\left(y+z\right)p^{2}+7p^{2}+16m^{2}\right)-3p^{4}+10\left(y+z\right)p^{4}\right)~ \Omega_{3} \left(\Delta_{2}\right)\nonumber \\
 & +\left(4p^{2}+8\left(y+z\right)p^{2}-32m^{2}\right)\frac{\tilde{p}_{\mu}\tilde{p}_{\nu}}{\tilde{p}^{2}}~ \Omega_{3}^{\mu \nu } \left(\Delta_{2}\right)\biggr)\nonumber \\
 & +\Gamma\left(4\right)\int d\varXi\biggl(p^{4}\left[-4m^{2}\left(z+w\right)+7m^{2}\left(z+w\right)^{2}+\left(z+w\right)^{2}p^{2}-4\left(z+w\right)^{3}p^{2}\right] ~ \Omega_{4} \left(\Delta_{3}\right)\nonumber \\
 & +\biggl[8\left(\left(z+w\right)^{2}p^{2}+4p^{2}-16\left(z+w\right)p^{2}-2m^{2}\bigg[8-14\left(z+w\right)\bigg]\right)p^{2}\frac{\tilde{p}_{\mu}\tilde{p}_{\nu}}{\tilde{p}^{2}}\nonumber \\
 & +\left(7m^{2}+p^{2}-12\left(z+w\right)p^{2}\right)p_{\mu}p_{\nu}\biggr]~ \Omega_{4}^{\mu \nu } \left(\Delta_{3}\right) +\frac{8}{\tilde{p}^{2}} p_{\mu}p_{\nu}\tilde{p}_{\lambda}\tilde{p}_{\beta}~ \Omega_{4}^{\mu \nu \lambda \beta} \left(\Delta_{3}\right) \biggr)\biggr\}.\label{eq:3.7}
\end{align}
Once again, the planar and nonplanar contributions can be computed
separately. For convenience, the expressions
for the planar and nonplanar contributions are written in
Appendix \ref{appD}, given by Eqs.\eqref{eq:3.8} and \eqref{eq:3.9},
respectively.

In contrast with the previous case, we see from Eqs.\eqref{eq:3.8}
and \eqref{eq:3.9}, that the planar and nonplanar parts do not sum
to zero at the commutative limit. This equation is a manifestation
of UV/IR mixing \footnote{Although, the \NC Maxwell-Chern-Simons theory in three dimensions is UV finite, we observe
explicitly a UV/IR mixing in our one-loop results.} \cite{minwalla}
\begin{align}
\left(\widetilde{\Pi}_{\mathrm{e}}\right)_{\mathrm{p}}\left(p^{2}\right)+\lim_{\theta\rightarrow~0}
\left(\widetilde{\Pi}_{\mathrm{e}}\right)_{\mathrm{n-p}}\left(p^{2}\right)=-\frac{e^{2}}{4\pi} \frac{1}{p^2}\left[\frac{1}{\left|\tilde{p}\right|}
+\frac{2m}{3}\right]\neq 0 . \label{eq:3.c2}
\end{align}

As it will be discussed later, one should already
notice that the presence of an UV/IR mixing term here might render
the theory to be inconsistent, spoiling hence the renormalizability
of the theory. Besides, it is worth notice that the UV/IR mixing in
$2+1$ dimensions appears in a less severe degree as
$\frac{1}{\left|\tilde{p}\right|}$, while in $3+1$ dimensions it is
given as $\frac{1}{\tilde{p}^2}$. Furthermore, in comparison to the
form factor $\Pi_{\mathrm{e}}$ outcome, we see that the commutative
limit for the contribution $\widetilde{\Pi}_{\mathrm{e}}$ is related
to the fact that it did not have a tree-level counterpart. So this
can be traced back to a purely \NC (quantum) effect.
%%%%%%%%%%%%%%%%%%%%%%%%%%%%%%%%%%%%%%%%%%%%%%%%%%%%%%%%%%%%%%%%%%%%%%%%%%%%%%%%%%%%%%%%%%%%%%%%%%%%%%%%%%%%%%%%%%%%%%%%%%%%%%%%%%%%%%%%
%%%%%%%%%%%%%%%%%%%%%%%%%%%%%%%%%%%%%%%%%%%%%%%%%%%%%%%%%%%%%%%%%%%%%%%%%%%%%%%%%%%%%%%%%%%%%%%%%%%%%%%%%%%%%%%%%%%%%%%%%%%%%%%%%%%%%%%%
\subsection{CP odd part $\Pi_{\mathrm{o}}^{\mathrm{\tiny{A}}}$}
\label{sec:4.3}

Moreover, we rewrite Eq.\eqref{eq:1.12} conveniently in terms of the quantities $ \Omega_{a}^{\mu \nu ...} \left(\Delta_{i}\right)$ such as
\begin{align}
\Pi_{\mathrm{o}}^{\mathrm{\tiny{A}}}\left(p^{2}\right) & =\frac{2mie^{2}}{p^{2}}\mu^{2\left(3-\omega\right)}\biggl\{5p^{2}\Gamma\left(2\right)\int d\Phi~ \Omega_{2}  \left(\Delta_{1}\right) \nonumber \\
 & +\Gamma\left(3\right)\int d\varUpsilon\left[\left(\left(-6+15x-5x^{2}\right)p^{2}+7m^{2}\right)p^{2} ~ \Omega_{3}  \left(\Delta_{2}\right) -5p_{\mu}p_{\nu} ~ \Omega_{3}^{\mu \nu } \left(\Delta_{2}\right) \right]\nonumber \\
 & -p^{2}\Gamma\left(4\right)\int d\varXi\left(z+w\right)^{2}\left\{ 2m^{2}-5\left(z+w\right)p^{2}+4p^{2}\right\} ~ \Omega_{4} \left(\Delta_{3}\right)\nonumber \\
 & -\Gamma\left(4\right)\int d\varXi\left(2m^{2}-5\left(z+w\right)p^{2}+4p^{2}-10\left(z+w\right)p^{2}\right)p_{\mu}p_{\nu}~ \Omega_{4}^{\mu \nu } \left(\Delta_{3}\right)\biggr\}.\label{eq:3.10}
\end{align}
Once again, we write down the expressions for the planar
and nonplanar contributions in Appendix \ref{appD}, explicitly
written in Eqs.\eqref{eq:3.11} and \eqref{eq:3.12}, respectively.

In agreement with our expectations, the sum of the planar and nonplanar contributions, Eqs.\eqref{eq:3.11} and \eqref{eq:3.12}, at the commutative limit, vanishes
\begin{align}
\left(\Pi_{\mathrm{o}}^{\mathrm{\tiny{A}}}\right)_{\mathrm{p}}\left(p^{2}\right)
+\lim_{\theta\rightarrow~0}\left(\Pi_{\mathrm{o}}^{\mathrm{\tiny{A}}}\right)_{\mathrm{n-p}}\left(p^{2}\right)=0. \label{eq:3.c3}
\end{align}
Exactly as it did happened with the form factor $\Pi_{\mathrm{e}}$
in \eqref{eq:3.c1}, this vanishing result is compatible
with the free nature of the commutative Maxwell-Chern-Simons theory,
where the form factor $\Pi_{\mathrm{o}}^{\mathrm{\tiny{A}}}$ has a tree-level  counterpart.

%%%%%%%%%%%%%%%%%%%%%%%%%%%%%%%%%%%%%%%%%%%%%%%%%%%%%%%%%%%%%%%%%%%%%%%%%%%%%%%%%%%%%%%%%%%%%%%%%%%%%%%%%%%%%%%%%%%%%%%%%%%%%%%%%%%%%%%%
%%%%%%%%%%%%%%%%%%%%%%%%%%%%%%%%%%%%%%%%%%%%%%%%%%%%%%%%%%%%%%%%%%%%%%%%%%%%%%%%%%%%%%%%%%%%%%%%%%%%%%%%%%%%%%%%%%%%%%%%%%%%%%%%%%%%%%%%
\subsection{NC odd part $\Pi_{\mathrm{o}}^{\mathrm{\tiny{S}}}$}
\label{sec:4.4}

Finally, we shall now analyze the NC odd part
$\Pi_{\mathrm{o}}^{\mathrm{\tiny{S}}}$; however, due to its
vanishing results, we present its final expressions here in order to
discuss its conclusion. We rewrite Eq.\eqref{eq:1.13} in terms of
the quantities $~ \Omega_{a}^{\mu \nu ...} \left(\Delta_{i}\right)$
such as
\begin{align}
\Pi_{\mathrm{o}}^{\mathrm{\tiny{S}}}\left(p^{2}\right) & =\frac{i}{\tilde{p}^{4}p^{2}}\mu^{2\left(3-\omega\right)}e^{2}u_{\mu}\tilde{p}_{\nu}\biggl\{4 \int d\Phi ~ \Omega_{2} ^{\mu \nu }  \left(\Delta_{1}\right)   \nonumber \\
 & +2\Gamma\left(3\right)\int d\varUpsilon\left(8m^{2}-p^{2}-2\left(y+z\right)p^{2}\right) ~ \Omega_{3}^{\mu \nu } \left(\Delta_{2}\right)  +2\Gamma\left(4\right)\int d\varXi \biggl( -2 p_{\lambda}p_{\beta}  ~ \Omega_{4}^{\mu \nu \lambda \beta} \left(\Delta_{3}\right)\nonumber \\
 & +p^2\left[4m^{2}-2\left(z+w\right)^{2}p^{2}-p^{2}-\left(z+w\right)\left(7m^{2}-4p^{2}\right)\right]
 ~ \Omega_{4}^{\mu \nu } \left(\Delta_{3}\right) \biggr)\biggr\}.\label{eq:3.13}
\end{align}
Based on the results for the momentum integration, Eqs.\eqref{eq:C.2} and \eqref{eq:C.4}, it is easy to conclude
that the planar contribution vanishes
\begin{align}
\left(\Pi_{\mathrm{o}}^{\mathrm{\tiny{S}}}\right)_{\mathrm{p}}\left(p^{2}\right) & =\frac{i}{2\tilde{p}^{4}p^{2}}  e^{2}u_{\mu}\tilde{p}_{\nu}\biggl\{\eta^{\mu\nu}A\left(\Delta_{1},\Delta_{2},\Delta_{3}\right)\nonumber \\
&+p_{\lambda}p_{\beta} \left(\eta^{\mu\nu}\eta^{\lambda\beta}+\eta^{\mu\lambda}\eta^{\nu\beta}+\eta^{\mu\beta}\eta^{\nu\lambda}\right)
B\left(\Delta_{1},\Delta_{2},\Delta_{3}\right)\biggr\} ,\nonumber\\
 & =0. \label{eq:3.14}
\end{align}
The last equality follows since $u.p=0$ and $u.\tilde{p}=0$. Now, proceeding in the same way, we have for the nonplanar part that
\begin{align}
\left(\Pi_{\mathrm{o}}^{\mathrm{\tiny{S}}}\right)_{\mathrm{n-p}}\left(p^{2}\right) & =-\frac{i}{2\tilde{p}^{4}p^{2}} e^{2}u_{\mu}\tilde{p}_{\nu}\biggl\{\frac{\eta^{\mu\nu}}{\omega}C\left(\Delta_{1},\Delta_{2}\right)+\frac{\tilde{p}^{\mu}\tilde{p}^{\nu}}{\tilde{p}^{2}}D\left(\Delta_{1},\Delta_{2}\right)\nonumber \\
&+p_{\lambda}p_{\beta}\biggl[\left(\eta^{\mu\nu}\eta^{\lambda\beta}+\eta^{\nu\lambda}\eta^{\mu\beta}+\eta^{\nu\beta}\eta^{\lambda\mu}\right)
E\left(\Delta_{3}\right)+\frac{\tilde{p}^{\lambda}\tilde{p}^{\beta}\tilde{p}^{\nu}\tilde{p}^{\mu}}{\tilde{p}^{4}}G\left(\Delta_{3}\right)\nonumber \\
 & +\left(\eta^{\lambda\beta}\frac{\tilde{p}^{\nu}\tilde{p}^{\mu}}{\tilde{p}^{2}}+\text{sym. permutations} \right)F\left(\Delta_{3}\right)\biggr]\biggr\}\nonumber\\
 & =0. \label{eq:3.15}
\end{align}

Again we have that the resulting expression is proportional to
$u.\tilde{p}=0$ and $u.p=0$. These vanishing results are in agreement
with the literature, since we can think about the Bose-Einstein
symmetry on the $\Pi^{\mu \nu}$, i.e. $\mu \leftrightarrow \nu$ and
$p\rightarrow -p$, in addition to its accidental symmetry $\theta
\rightarrow - \theta$, these combined facts show that the term $\Pi_{\mathrm{o}}^{\mathrm{\tiny{S}}}$ will
not be radiatively generated at higher order as well.
%%%%%%%%%%%%%%%%%%%%%%%%%%%%%%%%%%%%%%%%%%%%%%%%%%%%%%%%%%%%%%%%%%%%%%%%%%%%%%%%%%%%%%%%%%%%%%%%%%%%%%%%%%%%%%%%%%%%%%%%%%%%%%%%%%%%%%%%%%%%%%%%%%%%%%%%%%%%%%%%%%%%%%%%%%%%%%%%%%%%%%%%%%%%%%%%%%%%%%%%%%%%%%%%%%%%%%%%%%%%%%%%%%%%%%%%%%%%%%%%%%%%%%%%%%%%%%%%%%%%%%%%%%%%%%%%%%%%%%%%%%%%%%%%%%%%%%%%%%%%%%%%%%%%%%
\section{Dispersion relation and limiting cases}
\label{sec:5}

In order to establish some limits of special interest, we
consider the scaling $e A_{\mu} \rightarrow {\cal{A}}_{\mu}$ on the
Lagrangian \eqref{eq:1.1}, this implies into the following
change\footnote{ It is notable that the mass dimension of the gauge
field and the coupling constant in $2+1$ dimensions is equal to
$\frac{1}{2}$, hence the mass dimension of the new gauge field
${\cal{A}}_{\mu}$ and $\kappa$ is equal to 1 and 0, respectively.}
\begin{align}
\mathcal{L}=-\frac{1}{4e^{2}}{\cal{F}}_{\mu\nu} \star
{\cal{F}}^{\mu\nu}+\frac{\kappa}{2}\epsilon^{\mu\nu\lambda}\left({\cal{A}}_{\mu}
\star\partial_{\nu}{\cal{A}}_{\lambda}
+\frac{2}{3}{\cal{A}}_{\mu}\star {\cal{A}}_{\nu}\star
{\cal{A}}_{\lambda}\right) , \label{eq:4.1}
\end{align}
\noindent
 where ${\cal{F}}_{\mu\nu}=\partial_{\mu} {\cal{A}}_{\nu}-\partial_{\nu}
 {\cal{A}}_{\mu}+i[{\cal{A}}_{\mu},{\cal{A}}_{\nu}]_{\star}$ and also we have introduced a new parameter $\kappa e^{2}\equiv m$.
Therefore, from such Lagrangian we can immediately read three limits
of interest:
\begin{itemize}
\item (i) The NC Chern-Simons model is obtained when $e^{2}\rightarrow\infty$,
i.e. $m^{2}\rightarrow \infty$, so that the ratio $\kappa = m/e^{2}$ is kept finite;
\item (ii) The NC Maxwell model is obtained when $\kappa\rightarrow0$, i.e.
$m^{2}\rightarrow0$, so that $e^2$ is kept finite;
\item (iii) We can consider the low-momenta limit, also known as highly \NC limit, i.e. $p^{2}/m^{2} \rightarrow 0$ while $\tilde{p}$ is kept finite.
\end{itemize}
These three cases will be analyzed accordingly from our previous results in Sec. \ref{sec:4},
allowing us to obtain closed expression for the resulting form factors.

%%%%%%%%%%%%%%%%%%%%%%%%%%%%%%%%%%%%%%%%%%%%%%%%%%%%%%%%%%%%%%%%%%%%%%%%%%%%%%%%%%%%%%%%%%%%%%%%%%%%%%%%%%%%%%%%%%%%%%%%%%%%%%%%%%%%%%%%%%%%%%%%%%%%%%%%%%%%%%%%%%%%%%%%%%%%%%%%%%%%%%%%%%%%%%%%%%%%%%%%%%%%%%%%%%%%%%%%%%%%%%%%%%%%%%%%%%%%%%%%%%%%%%%%%%%%%%%%%%%%%%%%%%%%%%%%%%%%%%%%%%%%%%%%%%%%%%%%%%%%%%%%%%%%%%
\subsection{NC Chern-Simons model}
\label{sec:5.1}

This first limit is somehow laborious, and demands some careful
analysis. We can analyze the NC Chern-Simons theory by taking
directly the limit $m^{2}\rightarrow \infty$ in the expression
\eqref{eq:1.6}, we obtain the following result:
\begin{align}
\Pi_{\mu\nu}(p) & =\frac{m}{\kappa}\int\frac{d^{3}k}{(2\pi)^{3}}\frac{1}{k^{2}}\frac{1}{(p+k)^{2}}~\sin^{2}\left(\frac{p\wedge k}{2}
\right)~\left[k_{\mu}p{}_{\nu}-k_{\nu}p{}_{\mu}\right] \label{eq:4.2}
\end{align}
As usual, one can put the different poles under the same denominator,
which result into a change of the type: $Q_{\mu}=k_{\mu}+cp_{\mu}$,
where $c$ is some function of the Feynman parameter(s), for instance.
After this manipulation, one can easily find the expression
\begin{align}
\Pi_{\mu\nu}(p) & =\frac{m}{\kappa}\int\frac{d^{3}Q}{(2\pi)^{3}}\frac{1}{\left(Q^{2}-\Delta^{2}\right)^{2}}~\sin^{2}
\left(\frac{p\wedge Q}{2}\right)~\left[Q_{\mu}p{}_{\nu}-Q_{\nu}p{}_{\mu}\right]=0. \label{eq:4.3}
\end{align}

This vanishing result shows that no radiative correction to the gauge
field propagator is generated and accordingly its exhibits
the free nature of the \NC Chern-Simons theory, which is
in agreement with a previous analysis \cite{das}.

%%%%%%%%%%%%%%%%%%%%%%%%%%%%%%%%%%%%%%%%%%%%%%%%%%%%%%%%%%%%%%%%%%%%%%%%%%%%%%%%%%%%%%%%%%%%%%%%%%%%%%%%%%%%%%%%%%%%%%%%%%%%%%%%%%%%%%%%%%%%%%%%%%%%%%%%%%%%%%%%%%%%%%%%%%%%%%%%%%%%%%%%%%%%%%%%%%%%%%%%%%%%%%%%%%%%%%%%%%%%%%%%%%%%%%%%%%%%%%%%%%%%%%%%%%%%%%%%%%%%%%%%%%%%%%%%%%%%%%%%%%%%%%%%%%%%%%%%%%%%%%%%%%%%%%
\subsection{NC Maxwell model}

From either definition \eqref{eq:1.6} or form factors, Eqs.\eqref{eq:1.10}, \eqref{eq:1.11}, and \eqref{eq:1.12},
the limit $m\rightarrow0$ follows straightforwardly.
In this case, the form factor $\Pi_{\mathrm{e}}\left(p^{2}\right)$ is obtained from the sum of the planar and nonplanar contributions,
Eqs.\eqref{eq:3.5} and \eqref{eq:3.6}, respectively, and it yields to
\begin{align}
\Pi_{\mathrm{e}}\left(p^{2}\right) & =\frac{3e^{2}}{8\pi} \left|\tilde{p}\right| +\frac{e^{2}}{16\pi}\frac{1}{\left|p\right|}\biggl\{\frac{1}{2}\int \frac{d\varUpsilon}{\left(\Delta_{2}^{2}\right)^{\frac{3}{2}}}\biggl(\left(4\left(y+z\right)+2\right)\Delta_{2}^{2}~\Sigma^{\left(-\right)}\left(p, \tilde{p},\Delta_{2}\right)\nonumber \\
 & +\left(16\left(y+z\right)^{2}-10\left(y+z\right)+1\right)~\Sigma^{\left(+\right)}\left(p, \tilde{p},\Delta_{2}\right)  -16\Delta_{2}^{2} ~\Sigma\left(p, \tilde{p},\Delta_{2}\right)\biggr)\nonumber \\
 & -\frac{1}{4}\int \frac{d\varXi}{\left(\Delta_{3}^{2}\right)^{\frac{5}{2}}}\biggl(\left(z+w\right)^{2}\left(3-4\left(z+w\right)\right)~\Sigma^{\left(1\right)}\left(p, \tilde{p},\Delta_{3}\right)-\left(3-12\left(z+w\right)\right)\Delta_{3}^{2}~\Sigma^{\left(+\right)}\left(p, \tilde{p},\Delta_{3}\right)\nonumber \\
 & +\left(2+4\left(z+w\right)^{2}-8\left(z+w\right)\right)\Delta_{3}^{2}~\Sigma^{\left(2\right)}\left(p, \tilde{p},\Delta_{3}\right)-4\Delta_{3}^{4}~\Sigma^{\left(-\right)}\left(p, \tilde{p},\Delta_{3}\right)\biggr)\biggr\},\label{eq:4.4}
\end{align}
where in order to simplify the notation we have introduced the quantities
\begin{align}
\Sigma\left(p, \tilde{p},\Delta_{i}\right) & =1-e^{-\Delta_{i}\left|p\right|\left|\tilde{p}\right|}, \label{eq:4.5a} \\
\Sigma^{\left(\pm \right)}\left(p, \tilde{p},\Delta_{i}\right) & =1-\left(1\pm\Delta_{i}\left|p\right|\left|\tilde{p}\right|\right)e^{-\Delta_{i}\left|p\right|\left|\tilde{p}\right|}, \label{eq:4.5b}\\
\Sigma^{\left(1\right)}\left(p, \tilde{p},\Delta_{i}\right) & =3-\left(3+3\Delta_{i}\left|p\right|\left|\tilde{p}\right|+\Delta_{i}^{2}p^{2}\tilde{p}^{2}\right)e^{-\Delta_{i}\left|p\right|\left|\tilde{p}\right|}, \label{eq:4.5c}\\
\Sigma^{\left(2\right)}\left(p, \tilde{p},\Delta_{i}\right) & =1-\left[1+\Delta_{i}\left|p\right|\left|\tilde{p}\right|-\Delta_{i}^{2}p^{2}\tilde{p}^{2}\right]e^{-\Delta_{i}\left|p\right|\left|\tilde{p}\right|}, \label{eq:4.5d}
\end{align}
here the quantities $\Delta_{2}$ and $\Delta_{3}$ are given as
\begin{align}
\Delta_{2}^{2} & = -\left(y+z\right)\left(1-y-z\right),\\
\Delta_{3}^{2} & = -\left(z+w\right)\left(1-z-w\right).
\end{align}
Besides, from Eqs.\eqref{eq:3.8} and \eqref{eq:3.9}, we find for the NC transverse part $\widetilde{\Pi}_{\mathrm{e}}$ the following
\begin{align}
\widetilde{\Pi}_{\mathrm{e}}\left(p^{2}\right) & =\frac{e^{2}}{4\pi}\left[\left|\tilde{p}\right|-\frac{1}{\left|\tilde{p}\right|p^{2}}\right] \nonumber \\
& +\frac{e^{2}}{32\pi}\frac{1}{\left|p\right|}\biggl\{ \int \frac{d\varUpsilon}{\left(\Delta_{2}^{2}\right)^{\frac{3}{2}}}\biggl(\left(10\left(y+z\right)-3\right)~\Sigma^{\left(+\right)}\left(p, \tilde{p},\Delta_{2}\right)-4\left(1+2\left(y+z\right)\right)\Delta_{2}^{2}~\Sigma^{\left(-\right)}\left(p, \tilde{p},\Delta_{2}\right)\biggr)\nonumber \\
 & -\frac{1}{2}\int \frac{d\varXi}{\left(\Delta_{3}^{2}\right)^{\frac{5}{2}}}\biggl(\left(z+w\right)^{2}\left[1-4\left(z+w\right)\right]~\Sigma^{\left(1\right)}\left(p, \tilde{p},\Delta_{3}\right)+ \left(12\left(z+w\right)-1\right)\Delta_{3}^{2}~\Sigma^{\left(+\right)}\left(p, \tilde{p},\Delta_{3}\right)\nonumber \\
 & -4\left(1+2\left(z+w\right)^{2}-4\left(z+w\right)\right)\Delta_{3}^{2}~\Sigma^{\left(2\right)}\left(p, \tilde{p},\Delta_{3}\right)+8\Delta_{3}^{4}~\Sigma^{\left(-\right)}\left(p, \tilde{p},\Delta_{3}\right) \biggr)\biggr\}.\label{eq:4.6}
\end{align}
Finally, for the CP odd form factor $\Pi_{\mathrm{o}}^{\mathrm{\tiny{A}}}$, Eqs.\eqref{eq:3.11} and \eqref{eq:3.12}, we find as expected a vanishing result:
\begin{align}
\Pi_{\mathrm{o}}^{\mathrm{\tiny{A}}}\left(p^{2}\right) & =\left(\Pi_{\mathrm{o}}^{\mathrm{\tiny{A}}}\right)_{\mathrm{p}}\left(p^{2}\right)+\left(\Pi_{\mathrm{o}}^{\mathrm{\tiny{A}}}\right)_{\mathrm{n-p}}\left(p^{2}\right)=0. \label{eq:4.7}
\end{align}

%%%%%%%%%%%%%%%%%%%%%%%%%%%%%%%%%%%%%%%%%%%%%%%%%%%%%%%%%%%%%%%%%%%%%%%%%%%%%%%%%%%%%%%%%%%%%%%%%%%%%%%%%%%%%%%%%%%%%%%%%%%%%%%%%%%%%%%%%%%%%%%%%%%%%%%%%%%%%%%%%%%%%%%%%%%%%%%%%%%%%%%%%%%%%%%%%%%%%%%%%%%%
\subsubsection{Dispersion relation}

In the NC Maxwell theory we take the limit $m\rightarrow0$. Thus, we can write the complete propagator \eqref{eq:2.5} as the following
\begin{equation}
i\mathcal{D}_{\mu\nu}=\frac{1}{\left[p^{2}-\left|p\right|\Pi_{\mathrm{e}}^{\left(1\right)}
\left(p^{2}\right)\right]}\left(\eta_{\mu\nu}-\frac{p_{\mu}p_{\nu}}{p^{2}}-\frac{\tilde{p}_{\mu}\tilde{p}_{\nu}}{\tilde{p}^{2}}\right)
+\frac{1}{\left[p^{2}-\left|p\right|\widetilde{\Pi}_{\mathrm{e}}^{\left(1\right)}\left(p^{2}\right)\right]}\frac{\tilde{p}_{\mu}\tilde{p}_{\nu}}{\tilde{p}^{2}}+\frac{\xi}{p^{4}}p_{\mu}p_{\nu},
\label{eq:4.8}
\end{equation}
where the form factor expressions are defined so that $\left\{\Pi_{\mathrm{e}}^{\left(1\right)},\widetilde{\Pi}_{\mathrm{e}}^{\left(1\right)} \right\} = \left|p\right| \left\{\Pi_{\mathrm{e}} , \widetilde{\Pi}_{\mathrm{e}} \right\}$, with $\Pi_{\mathrm{e}} $ and $ \widetilde{\Pi}_{\mathrm{e}}$
given by Eqs.\eqref{eq:4.4} and \eqref{eq:4.6}, respectively.

In particular, the poles obtained above in Eq.\eqref{eq:4.8}, i.e.
$p^{2}-\left|p\right|\Pi_{e}^{\left(1\right)}\left(p^{2}\right)$,
reproduce a similar profile as those found on the three-dimensional
Yang-Mills theory \cite{pisarski,templeton}, which
allow us to (partially) discuss the infrared finiteness of the
model.

In order to illustrate the pole behavior, we can take small perturbations
around $p^{2}\tilde{p}^{2}$, so that at the leading-order the form factor expressions are reduced to
\begin{align}
\Pi_{\mathrm{e}}^{\left(1\right)}\left(p^{2}\right) & =\frac{7e^{2}}{24\pi}\left|p\right|
\left|\tilde{p}\right|-\frac{13ie^{2}}{512}p^{2}\tilde{p}^{2}, \label{eq:4.9a}
\end{align}
and
\begin{align}
\widetilde{\Pi}_{\mathrm{e}}^{\left(1\right)}\left(p^{2}\right) & =-\frac{e^{2}}{12\pi}\left|p\right|\left|\tilde{p}\right|
-\frac{e^{2}}{4\pi}\frac{1}{\left|p\right|\left|\tilde{p}\right|}+\frac{3ie^{2}}{256}p^{2}\tilde{p}^{2}.\label{eq:4.9b}
\end{align}
 These equations at the commutative limit change to the following:
\begin{align}
\lim_{\theta\rightarrow~0}\Pi_{\mathrm{e}}^{\left(1\right)}\left(p^{2}\right)&=0,\label{eq:new-1} \\
\lim_{\theta\rightarrow~0}\widetilde{\Pi}_{\mathrm{e}}^{\left(1\right)}\left(p^{2}\right)&=
-\frac{e^{2}}{4\pi}\frac{1}{\left|p\right|\left|\tilde{p}\right|}\label{eq:new-2}.
\end{align}
We note that \eqref{eq:new-1} is consistent with the commutative
Maxwell action, which is a free theory, however \eqref{eq:new-2},
similar to \eqref{eq:3.c2}, exhibits the UV/IR mixing effect in
$2+1$ dimensions and does not correspond to any counterpart term in
commutative Maxwell theory. The presence of the UV/IR
mixing in the NC Maxwell theory emphasize the fact that this theory
is not infrared finite.

%%%%%%%%%%%%%%%%%%%%%%%%%%%%%%%%%%%%%%%%%%%%%%%%%%%%%%%%%%%%%%%%%%%%%%%%%%%%%%%%%%%%%%%%%%%%%%%%%%%%%%%%%%%%%%%%%%%%%%%%%%%%%%%%%%%%%%%%%%%%%%%%%%%%%%%%%%%%%%%%%%%%%%%%%%%%%%%%%%%%%%%%%%%%%%%%%%%%%%%%%%%%%%%%%%%%%%%%%%%%%%%%%%%%%%%%%%%%%%%%%%%%%%%%%%%%%%%%%%%%%%%%%%%%%%%%%%%%%%%%%%%%%%%%%%%%%%%%%%%%%%%%%%%%%%
\subsection{Highly \NC Maxwell-Chern-Simons model}

We now study the third limiting case, which describes the low-momentum (or highly \NC)
behavior of NC Maxwell--Chern-Simons model.
For the highly \NC case, i.e. considering the limit $p^{2}/m^{2}\rightarrow0$ while
$\tilde{p}^2$ is kept finite, we can proceed in the exactly same way as before.
In this scenario, the complete form factor $\Pi_{\mathrm{e}}\left(p^{2}\right)$ is obtained
from Eqs.\eqref{eq:3.5} and \eqref{eq:3.6}, this results into
\begin{align}
\Pi_{\mathrm{e}}\left(p^{2}\right) & \simeq\frac{1}{16\pi \kappa}\biggl\{\int \frac{d\Phi}{\sqrt{\Delta_{1}^{2}}}\left[6+4\left(4-y\right)\frac{m^{2}}{p^{2}}\right]~\Sigma\left(m,\tilde{p},\Delta_{1}\right) \nonumber \\
 & +\frac{1}{2}\int\frac{ d\varUpsilon}{\left(\Delta_{2}^{2}\right)^{\frac{3}{2}}}\biggl(\left[2\left(13\left(y+z\right)-6\right)-16\frac{m^{2}}{p^{2}}\right]~\Sigma^{\left(+\right)}\left(m,\tilde{p},\Delta_{2}\right) \nonumber \\
 & +2\left(\left(2\left(y+z\right)+1\right)\Delta_{2}^{2}-8\frac{m^{2}}{p^{2}}\Delta_{2}^{2}+8\left(y+z\right)\left(1-y-z\right)\right)~\Sigma^{\left(-\right)}\left(m,\tilde{p},\Delta_{2}\right)\biggr)\nonumber  \\
 & -\frac{1}{4}\int \frac{d\varXi}{\left(\Delta_{3}^{2}\right)^{\frac{3}{2}}}\biggl(8\Sigma^{\left(+\right)}
 \left(m,\tilde{p},\Delta_{3}\right)-\left(8-14\left(z+w\right)\right)~\Sigma^{\left(2\right)}
 \left(m,\tilde{p},\Delta_{3}\right)-4\Delta_{3}^{2}~\Sigma^{\left(-\right)}\left(m,\tilde{p},\Delta_{3}\right)\biggr)\biggr\}
 \nonumber\\&+\mathcal{O}\left(\frac{p^{2}}{m^{2}}\right),\label{eq:4.10}
\end{align}
where the functions $\Sigma^{\left(i\right)}\left(m,\tilde{p},\Delta_{j}\right)$
are those defined before, Eqs.\eqref{eq:4.5a}--\eqref{eq:4.5d}, but now the quantities $\Delta _i$ are given by
\begin{align}
\Delta_{1}^{2}  =y , \quad \Delta_{2}^{2}  \simeq\left(x+z\right) +\mathcal{O}\left(\frac{p^{2}}{m^{2}}\right),\quad
\Delta_{3}^{2}  \simeq\left(y+w\right) +\mathcal{O}\left(\frac{p^{2}}{m^{2}}\right).
\end{align}
Moreover, we find for the NC transverse part $\widetilde{\Pi}_{\mathrm{e}}$,
the sum of the Eqs.\eqref{eq:3.8} and \eqref{eq:3.9}, the following expression
\begin{align}
\widetilde{\Pi}_{\mathrm{e}}\left(p^{2}\right) & \simeq \frac{1}{16\pi \kappa}  \biggl\{\int
\frac{d\Phi}{\sqrt{\Delta_{1}^{2}}}\biggl(2\left[2+\left(y-8\right)\frac{m^{2}}{p^{2}}\right]~\Sigma\left(m,\tilde{p},\Delta_{1}\right)
-\frac{4 m^{2}}{p^{2}}\left[\frac{\sqrt{y}}{m\left|\tilde{p}\right|}+y\right]e^{-\sqrt{y}m\left|\tilde{p}\right|}\biggr)\nonumber\\
 & +\frac{1}{2}\int \frac{d\varUpsilon}{\left(\Delta_{2}^{2}\right)^{\frac{3}{2}}}\biggl(\left[7+16\frac{m^{2}}{p^{2}}-30\left(y+z\right)\right]~\Sigma^{\left(+\right)}\left(m,\tilde{p},\Delta_{2}\right)\nonumber\\
 & -4\left[\left(1+2\left(y+z\right)-8\frac{m^{2}}{p^{2}}\right)\Delta_{2}^{2}+8\left(y+z\right)\left(1-y-z\right)\right]
 ~\Sigma^{\left(-\right)}\left(m,\tilde{p},\Delta_{2}\right)\biggr)\nonumber\\
 & -\frac{1}{4}\int \frac{d\varXi}{\left(\Delta_{3}^{2}\right)^{\frac{3}{2}}}\biggl(4\left(4-7\left(z+w\right)\right)
 ~\Sigma^{\left(2\right)}\left(m,\tilde{p},\Delta_{3}\right)-7\Sigma^{\left(+\right)}\left(m,\tilde{p},\Delta_{3}\right)+8\Delta_{3}^{2}
 ~\Sigma^{\left(-\right)}\left(m,\tilde{p},\Delta_{3}\right)\biggr)\biggr\}\nonumber\\&+\mathcal{O}\left(\frac{p^{2}}{m^{2}}\right), \label{eq:4.11}
\end{align}
At last, for the CP odd form factor $\Pi_{\mathrm{o}}^{\mathrm{\tiny{A}}}$, Eqs.\eqref{eq:3.11} and \eqref{eq:3.12}, we get
\begin{align}
\Pi_{\mathrm{o}}^{\mathrm{\tiny{A}}}\left(p^{2}\right) & \simeq-\frac{1}{8\pi}\frac{m}{\kappa}\biggl\{5\int \frac{d\Phi}{\sqrt{\Delta_{1}^{2}}}~\Sigma\left(m,\tilde{p},\Delta_{1}\right) -\frac{1}{2}\int \frac{d\varUpsilon}{\left(\Delta_{2}^{2}\right)^{\frac{3}{2}}}\left(7\Sigma^{\left(+\right)}\left(m,\tilde{p},\Delta_{2}\right)+5\Delta_{2}^{2}~\Sigma\left(m,\tilde{p},\Delta_{2}\right)\right) \nonumber\\
 &-\frac{1}{2}\int \frac{d\varXi}{\left(\Delta_{3}^{2}\right)^{\frac{3}{2}}}~\Sigma^{\left(+\right)}\left(m,\tilde{p},\Delta_{3}\right)\biggr\}+\mathcal{O}\left(\frac{p^{2}}{m^{2}}\right).\label{eq:4.12}
\end{align}

An important check of our results for the low-momenta limit is needed.
On one hand, in the commutative limit, Eqs.\eqref{eq:3.c1}, \eqref{eq:3.c2}, and \eqref{eq:3.c3}, we have
considered $\tilde{p}\rightarrow 0$, but this is explicitly read as $\theta \rightarrow 0$ when $p$ is kept finite.
On the other hand, however, we can equally consider $\tilde{p}\rightarrow 0$ as given
by $p\rightarrow 0$ with $\theta=\text{finite}$.
We can immediately conclude from the low-momenta limit expressions,
Eqs.\eqref{eq:4.10}, \eqref{eq:4.11}, and \eqref{eq:4.12}, that the latter
limit is in agreement with the (former) commutative limit.
Moreover, in order to understand this point consider the scale
$\theta=\frac{1}{\Lambda^{2}}$, hence, the commutative limit can be
interpreted as $\frac{p}{\Lambda}\ll 1$, where $p$ is the external momentum.
The condition $\frac{p}{\Lambda}\ll 1$ can then happen in two distinct cases:
\begin{equation}
  \left\{ \begin{aligned}
     \Lambda\rightarrow \infty,&  \quad  p=\text{finite} \\
      p\rightarrow 0,&  \quad \theta=\text{finite}
 \end{aligned}\right.
  \end{equation}
Hence, we see that these two limits are indeed the same and our
results are correct. In possess of the above explicit
results we can proceed to the analysis the dispersion relation
behavior for this case, and discuss the UV/IR mixing issue.
%%%%%%%%%%%%%%%%%%%%%%%%%%%%%%%%%%%%%%%%%%%%%%%%%%%%%%%%%%%%%%%%%%%%%%%%%%%%%%%%%%%%%%%%%%%%%%%%%%%%%%%%%%%%%%%%%%%%%%%%%%%%%%%%%%%%%%%%%%%%%%%%%%%%%%%%%%%%%%%%%%%%%%%%%%%%%%%%%%%%%%%%%%%%%%%%%%%%%%%%%%%%%%%%%%%%%%%%%%%%%%%%%%%%%%%%%%%%%%%%%%%%%%%%%%%%%%%%%%%%%%%%%%%%%%%%%%%%%%%%%%%%%%%%%%%%%%%%%%%%%%%%%%%%%%
\subsubsection{Dispersion relation}

The remaining integration on the Eqs.\eqref{eq:4.10}, \eqref{eq:4.11}, and \eqref{eq:4.12} can be computed analytically
without further complication thanks to the simplification due to the limit $p^{2}/m^{2}\rightarrow0$.
We, thus, obtain for the transverse form factor the explicit expression
\begin{align}
\Pi_{\mathrm{e}}\left(p^{2}\right) & \simeq     \frac{1}{2\pi\kappa}\frac{1}{m^{2}p^{2}\tilde{p}^{4}}\left[-48+6\sqrt{m^{2}\tilde{p}^{2}}+\left(m^{2}\tilde{p}^{2}\right)^{2}+\left(48+42\sqrt{m^{2}\tilde{p}^{2}}+18m^{2}\tilde{p}^{2}+5\left(m^{2}\tilde{p}^{2}\right)^{\frac{3}{2}}\right)e^{-\sqrt{m^{2}\tilde{p}^{2}}}\right] \nonumber \\
&+\frac{1}{120\pi \kappa}\frac{1}{\left(m^{2}\tilde{p}^{2}\right)^{3}}\biggl[39600+39720\sqrt{m^{2}\tilde{p}^{2}}+250m^{2}\tilde{p}^{2}+570\left(m^{2}\tilde{p}^{2}\right)^{\frac{3}{2}}\nonumber \\
&-240\left(m^{2}\tilde{p}^{2}\right)^{\frac{5}{2}}+131\left(m^{2}\tilde{p}^{2}\right)^{3} -5\biggl(7920+24\sqrt{m^{2}\tilde{p}^{2}}-2800m^{2}\tilde{p}^{2}\nonumber \\
 &-714\left(m^{2}\tilde{p}^{2}\right)^{\frac{3}{2}}+32\left(m^{2}\tilde{p}^{2}\right)^{2}+26\left(m^{2}\tilde{p}^{2}\right)^{\frac{5}{2}}+7\left(m^{2}\tilde{p}^{2}\right)^{3}\biggr)e^{-\sqrt{m^{2}\tilde{p}^{2}}}\biggr]+\mathcal{O} \left(\frac{p^{2}}{m^{2}}\right), \label{eq:4.13}
\end{align}
and next, for the NC transverse form factor, we find the result
\begin{align}
\widetilde{\Pi}_{\mathrm{e}}\left(p^{2}\right) & \simeq -\frac{1}{4\pi\kappa}\frac{1}{mp^{2}\tilde{p}^{3}}\left[-24+\left(m^{2}\tilde{p}^{2}\right)^{\frac{3}{2}}+\left(24+24\sqrt{m^{2}\tilde{p}^{2}}+13m^{2}\tilde{p}^{2}+4\left(m^{2}\tilde{p}^{2}\right)^{\frac{3}{2}}\right)e^{-\sqrt{m^{2}\tilde{p}^{2}}}\right] \nonumber \\
&-\frac{1}{240\pi \kappa}\frac{1}{\left(m^{2}\tilde{p}^{2}\right)^{\frac{5}{2}}}\biggl[-26880+3060m^{2}\tilde{p}^{2}+229\left(m^{2}\tilde{p}^{2}\right)^{\frac{3}{2}}-465\left(m^{2}\tilde{p}^{2}\right)^{2}\nonumber \\
 & +5\biggl(5376\left(1+\sqrt{m^{2}\tilde{p}^{2}}\right)+2076m^{2}\tilde{p}^{2}+284\left(m^{2}\tilde{p}^{2}\right)^{\frac{3}{2}}\nonumber \\
 &+11\left(m^{2}\tilde{p}^{2}\right)^{2}-10\left(m^{2}\tilde{p}^{2}\right)^{\frac{5}{2}}\biggr)
 e^{-\sqrt{m^{2}\tilde{p}^{2}}}\biggr]+\mathcal{O}\left(\frac{p^{2}}{m^{2}}\right).\label{eq:4.14}
\end{align}
Finally, for the CP odd form factor, we get
\begin{align}
\Pi_{\mathrm{o}}^{\mathrm{\tiny{A}}}\left(p^{2}\right) & \simeq-\frac{m}{24\pi \kappa}\frac{1}{\left(m^{2}\tilde{p}^{2}\right)^{\frac{3}{2}}}\biggl[2\left(3+9m^{2}\tilde{p}^{2}+\left(m^{2}\tilde{p}^{2}\right)^{\frac{3}{2}}\right)\nonumber \\
&-3\left(2+2\sqrt{m^{2}\tilde{p}^{2}}+7m^{2}\tilde{p}^{2}+7\left(m^{2}\tilde{p}^{2}\right)^{\frac{3}{2}}\right)e^{-\sqrt{m^{2}\tilde{p}^{2}}}\biggr]
+\mathcal{O}\left(\frac{p^{2}}{m^{2}}\right). \label{eq:4.15}
\end{align}

We can analyze the modifications caused into the dispersion relation by the noncommutativity
from examining first the formula \eqref{eq:2.6}.
By simplicity, we shall consider the contribution up to the lowest order in $\alpha = e^{2}/4\pi$.
This implies into the following expression for the renormalized mass
\begin{equation}
m_{\mathrm{ren}}\simeq m+\Pi^{(1)}
\end{equation}
where we have defined $\Pi^{(1)} =\Pi_{\mathrm{o}}^{\mathrm{\tiny{A}}}+m\Pi_{\mathrm{e}}
+\frac{m}{2}\widetilde{\Pi}_{\mathrm{e}}$, so that its expression reads
\begin{align}
\Pi^{(1)}& =  \frac{m}{8\pi\kappa}\frac{1}{m^{2}p^{2}\tilde{p}^{4}}\biggl[3\left(-64+16\sqrt{m^{2}\tilde{p}^{2}}+\left(m^{2}\tilde{p}^{2}\right)^{2}\right) \nonumber \\
 & +\left(192+144\sqrt{m^{2}\tilde{p}^{2}}+48\left(m^{2}\tilde{p}^{2}\right)+7\left(m^{2}\tilde{p}^{2}\right)^{\frac{3}{2}}-4\left(m^{2}\tilde{p}^{2}\right)^{2}\right)e^{-\sqrt{m^{2}\tilde{p}^{2}}}\biggr]    \nonumber \\
 &+\frac{m}{96\pi\kappa}\frac{1}{\left(m^{2}\tilde{p}^{2}\right)^{3}}\biggl[-31680+5280\sqrt{m^{2}\tilde{p}^{2}}+2016m^{2}\tilde{p}^{2}-180\left(m^{2}\tilde{p}^{2}\right)^{\frac{3}{2}}\nonumber \\
 & -171\left(m^{2}\tilde{p}^{2}\right)^{\frac{5}{2}}+51\left(m^{2}\tilde{p}^{2}\right)^{3} +\biggl(31680+26400\sqrt{m^{2}\tilde{p}^{2}}+8544m^{2}\tilde{p}^{2}+804\left(m^{2}\tilde{p}^{2}\right)^{\frac{3}{2}}\nonumber \\
 &-388\left(m^{2}\tilde{p}^{2}\right)^{2}-31\left(m^{2}\tilde{p}^{2}\right)^{\frac{5}{2}}+66\left(m^{2}\tilde{p}^{2}\right)^{3}\biggr)e^{-\sqrt{m^{2}\tilde{p}^{2}}}\biggr]+\mathcal{O}\left(\frac{\left|p\right|}{m \kappa}\right)+\mathcal{O}\left(\frac{m \left|\tilde{p}\right|}{\kappa}\right).
\end{align}
In order to illustrate our result, we can consider small perturbation in powers of $m^{2}\tilde{p}^{2}$, and we find at the leading-order that
\begin{align}
\Pi^{(1)} & = -\frac{1}{8\pi\kappa}\frac{m^{2}}{p^{2}\left|\tilde{p}\right|}
+\frac{m}{\kappa}\left[-\frac{317}{1260\pi}\sqrt{m^{2}\tilde{p}^{2}}
+....\right]+\mathcal{O}\left(\frac{\left|p\right|}{m
\kappa}\right)+\mathcal{O}\left(\frac{m^2
\tilde{p}^2}{\kappa^2}\right) +\mathcal{O}\left(\frac{m^{3}\left|\tilde{p}\right|}{\kappa
p^{2}}\right). \label{eq:4.16}
\end{align}
Finally, substituting \eqref{eq:4.16} into \eqref{eq:2.8}, we obtain the following dispersion relation
\begin{equation}
\omega^{2}=\vec{p}^{2}+m^{2} -\frac{1}{4\pi\kappa}\frac{m^{3}}{p^{2}\left|\tilde{p}\right|}
 +\mathcal{O}\left(\frac{\left|p\right|}{m \kappa}\right)+\mathcal{O}\left(\frac{m
\left|\tilde{p}\right|}{\kappa}\right) +\mathcal{O}\left(\frac{m^{3}\left|\tilde{p}\right|}{\kappa
p^{2}}\right) .
\end{equation}
In particular, we see from this expression that the
highly \NC behavior has an UV/IR instability in the NC momentum and
consequently the theory is not infrared finite. These facts can be
seen as originating from the general result in Eq.\eqref{eq:3.c2}.

Furthermore, the insertion of the form factor expressions
\eqref{eq:4.13}, \eqref{eq:4.14}, and \eqref{eq:4.15} into the
relation \eqref{eq:2.5} gives us the one-loop photon propagator in
the low-energy limit of the noncommutative Maxwell-Chern-Simons
theory. One of the main physical consequences of this corrected
propagator is that we can determine the \NC one-loop corrections to
the electrostatic potential energy of the Maxwell-Chern-Simons
model. By taking into account the time-like photons from the above
result, we find the expression \footnote{Here, by means of clarity,
we have restored the usual notation in terms of $e^2 =m/\kappa$.}
\begin{equation}
V^{1-loop}\left(r\right)=e^{2}\int dx_{0}\int\frac{d^{3}p}{\left(2\pi\right)^{3}}\frac{1}{p^{2}-m_{ren}^{2}}e^{-i\vec{p}.\vec{r}+ip_{0}x_{0}}\left[1+\Pi_{e}\right],
\end{equation}
that by setting $\Pi_{e}=0$, the free part of the
potential is given by $V^{free}\left(r\right)
=-\frac{e^{2}}{2\pi}K_{0}\left(mr\right)
\rightarrow\frac{e^{2}}{2\pi} \ln\left(mr\right)$ as $mr\ll1 $. Now,
considering only the \NC contribution, i.e. by taking into account
the leading contribution in Eqs.\eqref{eq:4.13} and \eqref{eq:4.16},
we obtain the following deviation for the potential
\begin{align}
\delta V^{1-loop}\left(r\right) &=-\frac{\left|\theta\right|e^{4}}{20160\pi^{2}}\int p^{3}dp\frac{1}{p^{3}\left(p^{2}+m^{2}\right)-\frac{e^{2}}{4\pi}\frac{m^{2}}{\left|\theta\right|}}\left[4009p^{2}+4284m^{2}\right]J_{0}\left(pr\right), \nnr
& =-\frac{a\left|\theta\right|e^{4}}{\pi^{2}}\frac{1}{r}\int_{0}^{\infty}dp\frac{p^{3}\left(p^{2}+
b m^{2}r^2 \right)}{p^{3}\left(p^{2}+m^{2}r^2 \right)-\frac{e^{2}}{4\pi}\frac{m^{2}r^5}{\left|\theta\right|}}J_{0}\left(p \right),
\end{align}
in which $a=0.198$ and $b=1.068$. By means of a simple analysis, we see that the denominator of the
above integrand has one real positive root and four complex roots,
named here as $p_{0}$ and $p_{j}$, respectively. Due to its pole structure,
the integrand can be written as a pseudo function plus a delta function. Hence, after some computation \cite{Gradshteyn}, we finally arrive at
\begin{align}
\delta V^{1-loop}\left(r\right) &=-\frac{a\left|\theta\right|e^{4}}{\pi^{2}}\frac{1}{r}\bigg\{
-i\pi~\frac{A(p_{0})}{B(p_{0})}J_{0}(p_{0})+ 1 \nnr
&+\sum\limits_{j=1}^{4}c_{j}\bigg[K_{0}(-i p_{j})+\frac{\pi}{2}\bigg(iJ_{0}( p_{j})-H_{0}( p_{j})\bigg)\bigg]
\bigg\},
\end{align}
where $A=p_{0}^{3}\left(p_{0}^{2}+ b m^{2}r^2\right)$,
$B=p_{0}^{3}\left(p_{0}^{2}+m^{2}r^2
\right)-\frac{e^{2}}{4\pi}\frac{m^{2}r^5}{\left|\theta\right|} $ and
$H$ is the Struve function. Also, the coefficient $c_{j}$ arises
from the fraction decomposition which is given by
\begin{eqnarray}
\frac{A(p)}{B(p)}=1+\sum\limits_{j=o}^{4}\frac{c_{j}}{p-p_{j}}
\end{eqnarray}
with $c_{j}=\frac{A(p_{j})}{B'(p_{j})}$. Furthermore, by means of
illustration, we consider the behavior of the above expression again
at $mr\ll1 $, so that we can compare it with the usual free result.
Hence, by using the asymptotic expansion of the Bessel and Struve
functions \cite{Gradshteyn}, we find
\begin{equation}
\delta V^{1-loop}\left(r\right) =-\frac{a\left|\theta\right|e^{4}}{\pi^{2}}\frac{1}{r}\left(1
-i\pi~\frac{A(p_{0})}{B(p_{0})} \right).
\end{equation}
It is important to emphasize the strong departure due to the
noncommutativity of the leading radial dependence of the above
expression, $\delta V^{1-loop}  \propto \frac{1}{r}$, when compared
to the usual Maxwell-Chern-Simons (confining) static potential
energy $V^{free}  \propto \ln r$, at $mr\ll1 $.

It is worth noticing that an investigation of the Lamb shift effect
in \NC $QED_{4}$ was carried out in \cite{Masud}. Since a complete
discussion for this physical process requires us to take into account
the charge renormalization, this will be considered further in the
forthcoming paper \cite{ghasemkhani}.

%%%%%%%%%%%%%%%%%%%%%%%%%%%%%%%%%%%%%%%%%%%%%%%%%%%%%%%%%%%%%%%%%%%%%%%%%%%%%%%%%%%%%%%%%%%%%%%%%%%%%%%%%%%%%%%%%%%%%%%%%%%%%%%%%%%%%%%%%%%%%%%%%%%%%%%%%%%%%%%%%%%%%%%%%%%%%%%%%%%%%%%%%%%%%%%%%%%%%%%%%%%%%%%%%%%%%%%%%%%%%%%%%%%%%%%%%%%%%%%%%%%%%%%%%%%%%%%%%%%%%%%%%%%%%%%%%%%%%%%%%%%%%%%%%%%%%%%%%%%%%%%%%%%%%%
\section{Concluding remarks}
\label{sec:6}

In this paper, we have studied in complete detail the gauge field
complete propagator at one-loop order in the NC Maxwell-Chern-Simons theory.
A careful account covering all the renormalizability aspects
of this two-point function has been presented, in particular by establishing the
respective renormalization constants and subsequently the gauge field renormalized mass.
It is worth mentioning that, as expected from a gauge theory, a multiplicative
renormalization holds for the theory.

We first discussed in detail the tensor structure of the gauge field
self-energy at one-loop order. This has been supplemented by a full
account on the discrete symmetries for a three-dimensional \NC spacetime.
The explicit expressions of the form factor were calculated by following the standard rules of Feynman integration.
In particular, we found that the commutative limit of the complete
form factor $\widetilde{\Pi}_{\mathrm{e}}$ displays a manifestation
of IR/UV mixing, since the planar and nonplanar contributions sum to a nonvanishing result.
Besides, we explicitly showed that the NC $\mathbf{CP}$-odd form factor
$\Pi_{\mathrm{o}}^{\mathrm{\tiny{S}}}$ identically vanishes.

In order to discuss some physical consequences of the considered
model, we have scrutinized some particular limits: (i) the NC
Chern-Simons theory; (ii) the NC Maxwell theory; and (iii) the
low-momenta limit, highly \NC Maxwell-Chern-Simons theory. First, we
found that, as expected, the NC Chern-Simons theory is actually a free
theory. Next, we showed that the massless limit, $m=0$,
is well behaved in this context, and that the dispersion relation
for the NC Maxwell theory displays the same profile as in the
three-dimensional Yang-Mills theory, with no radiative mass
generation, but with an UV/IR mixing instability.

Finally, the highly \NC limit was considered, and
analytical expressions have been obtained for the form factors.
Within this context, we examined its dispersion relation and found
that it is not infrared finite, more precisely an UV/IR mixing
instability due to the NC momentum. Besides, using the
one-loop expressions of the form factors, we have determined the \NC
corrections to the photon propagator in the low-momenta limit. As a
physical outcome of the one-loop gauge field propagator, we have
discussed the noncommutative corrections to the electrostatic
potential. In particular, the low-momenta limit of the
Maxwell-Chern-Simons theory (when coupled to matter fields) is of
major interest for physical application in planar materials, in
particular to the description of new materials in the framework of
condensed matter physics
\cite{Miransky:2015ava,Martin-Ruiz:2015skg}, which allows the use of
effective low-energy models.

It is worth mentioning that a complete account of the
\NC Maxwell-Chern-Simons theory renormalizability must contain an
analysis of the vertex functions. This complementary study is now
under scrutiny \cite{ghasemkhani}. The analysis takes into account
the renormalization of the 3-point vertex function and ghost sector
, as well other vertex functions, allowing us then to fully
discuss the renormalization of the gauge coupling, determining
the theory's beta function, as well as the presence of the UV/IR
mixing and infrared finiteness.

%%%%%%%%%%%%%%%%%%%%%%%%%%%%%%%%%%%%%%%%%%%%%%%%%%%%%%%%%%%%%%%%%%%%%%%%%%%%%%%%%%%%%%%%%%%%%%%%
\subsection*{Acknowledgments}
The authors would like to thank the anonymous referee for his/her comments
and suggestions to improve this paper. We would like to express our special gratitude to M.M. Sheikh-Jabbari for his valuable comments and enlightening discussions. Also, the useful remarks from M. Khorrami are greatly appreciated. R.B. thankfully acknowledges FAPESP for support, Project No. 2011/20653-3.
\appendix
%%%%%%%%%%%%%%%%%%%%%%%%%%%%%%%%%%%%%%%%%%%%%%%%%%%%%%%%%%%%%%%%%%%%%%%%%%%%%%%%%%%%%%%%%%%%%%%%%%%%%%%%%%%%%%%%%%%%%%%%%%%%%%%%%%%%%%%%%%%%%%%%%%%%%%%%%%%%%%%%%%%%%%%%%%%%%%%%%%%%%%%%%%%%%%%%%%%%%%%%%%%%%%%%%%%%%%%%%%%%%%%%%%%%%%%%%%%%%%%%%%%%%%%%%%%%%%%%%%%%%%%%%%%%%%%%%%%%%%%%%%%%%%%%%%%%%%%%%%%%%%%%%%%%%%
\section{One-loop analysis of the photon self energy}
\label{appA}

We shall now write down the full contribution one-loop expression
for the photon self-energy; see Figure \ref{oneloopdiagrams}. This contribution has the following form,
\begin{align}
\Pi_{\mu\nu} =\Pi_{\mu\nu}^{\mathrm{g}}+\Pi_{\mu\nu}^{\mathrm{gh}}+\Pi_{\mu\nu}^{\mathrm{t}},\label{eq:A.1}
\end{align}
where the explicit expression for the the ghost, cubic, and quartic
self-interacting diagrams are given by
\begin{align}
\Pi_{\mu\nu}^{\mathrm{gh}}(p) & =2e^{2}\int\frac{d^{3}k}{(2\pi)^{3}}\frac{1}{k^{2}}\frac{1}{(p+k)^{2}}~\sin^{2}\left(\frac{p\wedge k}{2}\right)~k_{\mu}(k+p)_{\nu},\label{eq:A.2a}\\
\Pi_{\mu\nu}^{\mathrm{g}}(p) & =e^{2}\int\frac{d^{3}k}{(2\pi)^{3}}~\sin^{2}\left(\frac{p\wedge k}{2}\right)~\frac{1}{k^{2}(k^{2}-m^{2})}~\frac{1}{(p+k)^{2}[(p+k)^{2}-m^{2}]}~{\cal {N}}_{\mu\nu}^{\mathrm{g}},\label{eq:A.2b}\\
\Pi_{\mu\nu}^{\mathrm{t}}(p) & =2e^{2}\int\frac{d^{3}k}{(2\pi)^{3}}~~\sin^{2}\left(\frac{p\wedge k}{2}\right)~\frac{\eta_{\mu\nu}k^{2}+k_{\mu}k_{\nu}}{k^{2}(k^{2}-m^{2})},\label{eq:A.2c}
\end{align}
where the tensor at the numerator of Eq.\eqref{eq:A.2b} is defined
as
\begin{align}
{\cal N}_{\mu\nu}^{\mathrm{g}} & =\left(im\epsilon_{\mu\alpha\beta}+(p+2k)_{\mu}\eta_{\alpha\beta}+(p-k)_{\beta}\eta_{\mu\alpha}-(2p+k)_{\alpha}\eta_{\mu\beta}\right)\nonumber \\
 & \times\left(im\epsilon_{\nu\rho\sigma}-(p+2k)_{\nu}\eta_{\rho\sigma}+(k-p)_{\sigma}\eta_{\rho\nu}+(2p+k)_{\rho}\eta_{\nu\sigma}\right)\nonumber \\
 & \times\left(k^{2}\eta^{\alpha\rho}-k^{\alpha}k^{\rho}+im\epsilon^{\alpha\rho\lambda}k_{\lambda}\right)\left((p+k)^{2}\eta^{\beta\sigma}-(p+k)^{\beta}(p+k)^{\sigma}-im\epsilon^{\beta\sigma\xi}(p+k)_{\xi}\right).\label{eq:A.3}
\end{align}
We note that the denominator on these three contributions is different.
Hence, we can write the complete contribution in the following way
\begin{align}
\Pi_{\mu\nu}\left(p\right)=e^{2}\int\frac{d^{3}k}{(2\pi)^{3}}~\sin^{2}\left(\frac{p\wedge k}{2}\right)~\frac{{\cal N}_{\mu\nu}^{\mathrm{g}}+2{\cal N}_{\mu\nu}^{\mathrm{gh}}+2{\cal N}_{\mu\nu}^{\mathrm{t}}}{k^{2}(k^{2}-m^{2})(p+k)^{2}((p+k)^{2}-m^{2})}\label{eq:A.4}
\end{align}
where we have conveniently introduced the new tensor quantities:
\begin{align}
{\cal N}_{\mu\nu}^{\mathrm{gh}} & =m^{4}\left(k_{\mu}k_{\nu}+k_{\mu}p_{\nu}\right)-m^{2}\left(2k^{2}+p^{2}+2p.k\right)\left(k_{\mu}k_{\nu}+k_{\mu}p_{\nu}\right)\nonumber \\
 & +k^{2}\left(k^{2}+2p.k+p^{2}\right)\left(k_{\mu}k_{\nu}+k_{\mu}p_{\nu}\right),\label{eq:A.5}\\
{\cal N}_{\mu\nu}^{\mathrm{t}} & =-m^{2}\left(k^{2}+2p.k+p^{2}\right)\left(k^{2}\eta_{\mu\nu}+k_{\mu}k_{\nu}\right)\nonumber \\
 & +\left(k^{4}+2k^{2}p^{2}+p^{4}+4k^{2}(p.k)+4p^{2}(p.k)+4(p.k)^{2}\right)\left(k^{2}\eta_{\mu\nu}+k_{\mu}k_{\nu}\right).\label{eq:A.6}
\end{align}

%%%%%%%%%%%%%%%%%%%%%%%%%%%%%%%%%%%%%%%%%%%%%%%%%%%%%%%%%%%%%%%%%%%%%%%%%%%%%%%%%%%%%%%%%%%%%%%%%%%%%%%%%%%%%%%%%%%%%%%%%%%%%%%%%%%%%%%%%%%%%%%%%%%%%%%%%%%%%%%%%%%%%%%%%%%%%%%%%%%%%%%%%%%%%%%%%%%%%%%%%%%%%%%%%%%%%%
\section{Tensor structure of the photon self-energy}
\label{appB}

In order to discuss the tensor structure of the complete photon propagator,
we shall introduce and consider the following vectors, $p_{\mu}$,
$\tilde{p}_{\mu}=\theta_{\mu\nu}p^{\nu}$ and $u_{\mu}=\epsilon_{\mu\alpha\beta}p^{\alpha}\tilde{p}^{\beta}$
, as an orthogonal basis. In particular, it is easy to see that
\begin{align}
p_{\mu}\tilde{p}^{\mu}=0,\quad p^{\mu}u_{\mu}=0,\quad\tilde{p}^{\mu}u_{\mu}=0\label{eq:B.1}
\end{align}
and that the completeness relation is also satisfied:
\begin{align}
\frac{u_{\mu}u_{\nu}}{u^{2}}+\frac{\tilde{p}_{\mu}\tilde{p}_{\nu}}{\tilde{p}^{2}}+\frac{p_{\mu}p_{\nu}}{p^{2}}=\eta_{\mu\nu}.\label{eq:B.2}
\end{align}
The polarization tensor in this basis is written as
\begin{equation}
\Pi_{\mu\nu}=(a_{1}p_{\mu}+a_{2}\tilde{p}_{\mu}+a_{3}u_{\mu})p_{\nu}+(b_{1}p_{\mu}+b_{2}\tilde{p}_{\mu}+b_{3}u_{\mu})\tilde{p}_{\nu}
+(c_{1}p_{\mu}+c_{2}\tilde{p}_{\mu}+c_{3}u_{\mu})u_{\nu}.
\end{equation}
Applying the Ward identity $p^{\mu}\Pi_{\mu\nu}=0$ and $p^{\nu}\Pi_{\mu\nu}=0$, this directly leads to $a_{1}=a_{2}=a_{3}=b_{1}=c_{1}=0$.
Hence, we are left with the expression
\begin{align}
\Pi_{\mu\nu}&=(b_{2}\tilde{p}_{\mu}+b_{3}u_{\mu})\tilde{p}_{\nu}
+(c_{2}\tilde{p}_{\mu}+c_{3}u_{\mu})u_{\nu}\nonumber\\
&=b_{2}\tilde{p}_{\mu}\tilde{p}_{\nu}+d_{1}(u_{\mu}\tilde{p}_{\nu}+u_{\nu}\tilde{p}_{\mu})
+d_{2}(\tilde{p}_{\mu}u_{\nu}-\tilde{p}_{\nu}u_{\mu})+c_{3}u_{\mu}u_{\nu},
\end{align}
in which, by convenience, we have rewritten the terms
$b_{3}u_{\mu}\tilde{p}_{\nu}+c_{2}\tilde{p}_{\mu}u_{\nu}$ as
symmetric and antisymmetric parts. Furthermore, it is easy to show
that the antisymmetric part can be revised as
\begin{align}
\tilde{p}_{\mu}u_{\nu}-\tilde{p}_{\nu}u_{\mu}=\epsilon_{\mu\nu\lambda}p^{\lambda}\tilde{p}^{2}.
\end{align}
Using this result and also the completeness relation \eqref{eq:B.2},
we can write the photon self-energy in a clear and appropriated form; see \eqref{eq:B.4b}.
 Now, at this step, we construct the general form for the $1PI$ function $\Gamma_{\mu\nu}$
 for the NC Maxwell-Chern-Simons theory using the defined basis
\begin{align}
\Gamma^{\mu\nu}=\Gamma_{\mathrm{\tiny{tree-level}}}^{\mu\nu}+\Pi_{\mathrm{\tiny{loop-level}}}^{\mu\nu}\label{eq:B.3}
\end{align}
where the $1PI$ two-point function and the polarization tensor are
, respectively, given by
\begin{align}
\Gamma_{\mathrm{\tiny{tree-level}}}^{\mu\nu} & =-p^{2}\eta^{\mu\nu}+(1-\frac{1}{\xi})p^{\mu}p^{\nu}+im\epsilon^{\mu\nu\lambda}p_{\lambda}\label{eq:B.4a}\\
\Pi_{\mathrm{\tiny{loop-level}}}^{\mu\nu} & =\bigg(\eta^{\mu\nu}-\frac{p^{\mu}p^{\nu}}{p^{2}}\bigg)
\Pi^{\star}_{\mathrm{e}}+\frac{\tilde{p}^{\mu}\tilde{p}^{\nu}}{\tilde{p}^{2}}~
\widetilde{\Pi}^{\star}_{\mathrm{e}}+i\epsilon^{\mu\nu\lambda}p_{\lambda}
\Pi_{\mathrm{o}}^{\mathrm{\tiny{A}}}+\bigg(\tilde{p}^{\mu}u^{\nu}+\tilde{p}^{\nu}u^{\mu}\bigg)\Pi_{\mathrm{o}}^{\mathrm{\tiny{S}}}.\label{eq:B.4b}
\end{align}
with which the tensor structure is in agreement \cite{caporaso}. We
can obtain the complete propagator expression by means of the
standard functional relation
$\Gamma^{\lambda\mu}\mathcal{D}_{\mu\nu}=i\delta_{\nu}^{\lambda}$.
After some laborious calculation, we find that the complete
propagator has the following general expression,
\begin{align}
i\mathcal{D}_{\mu\nu} & =\frac{p^{2}-\Pi^{\star}_{\mathrm{e}}-\widetilde{\Pi}^{\star}_{\mathrm{e}}}{\mathcal{R}}\eta_{\mu\nu}+\left(\frac{-p^{2}+\Pi^{\star}_{\mathrm{e}}+\widetilde{\Pi}^{\star}_{\mathrm{e}}}{\mathcal{R}}+\frac{\xi}{p^{2}}\right)\frac{p_{\mu}p_{\nu}}{p^{2}}+\frac{\widetilde{\Pi}^{\star}_{\mathrm{e}}}{\mathcal{R}}\frac{\tilde{p}_{\mu}\tilde{p}_{\nu}}{\tilde{p}^{2}}\nonumber \\
 & +\frac{\Pi_{\mathrm{o}}^{\mathrm{\tiny{S}}}}{\mathcal{R}}\left(\tilde{p}_{\mu}u_{\nu}+u_{\mu}\tilde{p}_{\nu}\right)+\frac{m+\Pi_{\mathrm{o}}^{\mathrm{\tiny{A}}}}{\mathcal{R}}i\varepsilon_{\mu\nu\lambda}p^{\lambda},\label{eq:B.5}
\end{align}
where $\mathcal{R}=(p^{2}-\Pi^{\star}_{\mathrm{e}})(p^{2}-\Pi^{\star}_{\mathrm{e}}-\widetilde{\Pi}^{\star}_{\mathrm{e}})+p^{2}\bigg[(\tilde{p}^{2}\Pi_{\mathrm{o}}^{\mathrm{\tiny{S}}})^{2}-(m+\Pi_{\mathrm{o}}^{\mathrm{\tiny{A}}})^{2}\bigg]$.

In order to conclude this discussion, we shall now determine the coefficients
appearing in the expression \eqref{eq:B.4b} for the $1PI$ form factors
$\Pi^{\star}_{\mathrm{e}}$, $\widetilde{\Pi}^{\star}_{\mathrm{e}}$, $\Pi_{\mathrm{o}}^{\mathrm{\tiny{A}}}$
and $\Pi_{\mathrm{o}}^{\mathrm{\tiny{S}}}$. These are found from
the following identities:
\begin{align}
\Pi^{\star}_{\mathrm{e}}= & \eta_{\mu\nu}\Pi^{\mu\nu}-\frac{\tilde{p}_{\mu}\tilde{p}_{\nu}}{\tilde{p}^{2}}\Pi^{\mu\nu},\label{eq:B.6a}\\
\widetilde{\Pi}^{\star}_{\mathrm{e}}= & -\eta_{\mu\nu}\Pi^{\mu\nu}+2\frac{\tilde{p}_{\mu}\tilde{p}_{\nu}}{\tilde{p}^{2}}\Pi^{\mu\nu},\label{eq:B.6b}\\
\Pi_{\mathrm{o}}^{\mathrm{\tiny{A}}}= & \frac{i}{2p^{2}}\epsilon_{\mu\nu\alpha}p^{\alpha}\Pi^{\mu\nu},\label{eq:B.6c}\\
\Pi_{\mathrm{o}}^{\mathrm{\tiny{S}}}= & -\frac{1}{2\tilde{p}^{4}p^{2}}\left(u_{\mu}\tilde{p}_{\nu}+u_{\nu}\tilde{p}_{\mu}\right)\Pi^{\mu\nu}.\label{eq:B.6d}
\end{align}

%%%%%%%%%%%%%%%%%%%%%%%%%%%%%%%%%%%%%%%%%%%%%%%%%%%%%%%%%%%%%%%%%%%%%%%%%%%%%%%%%%%%%%%%%%%%%%%%%%%%%%%%%%%%%%%%%%%%%%%%%%%%%%%%%%%%%%%%%%%%%%%%%%%%%%%%%%%%%%%%%%%%%%%%%%%%%%%%%%%%%%%%%%%%%%%%%%%%%%%%%%%%%%%%%%%%%%
\section{Nonplanar integrals}
\label{appC}

Throughout the paper, we have made use of some known results involving momentum
integration. We shall recall some of these results, in particular
those involving a nonplanar factor. The simplest integration reads
\begin{equation}
\int\frac{d^{\omega}q}{\left(2\pi\right)^{\omega}}\frac{1}{\left(q^{2}-s^{2}\right)^{a}}e^{ik\wedge q}=\frac{2i\left(-\right)^{a}}{\left(4\pi\right)^{\frac{\omega}{2}}}\frac{1}{\Gamma\left(a\right)}\frac{1}{\left(s^{2}\right)^{a-\frac{\omega}{2}}}\left(\frac{\left|\tilde{k}\right|s}{2}\right)^{a-\frac{\omega}{2}}K_{a-\frac{\omega}{2}}\left(\left|\tilde{k}\right|s\right).\label{eq:C.1}
\end{equation}
Next, we have the integration
\begin{align}
\int\frac{d^{\omega}q}{\left(2\pi\right)^{\omega}}\frac{q^{\mu}q^{\nu}}{\left(q^{2}-s^{2}\right)^{a}}e^{ik\wedge q} & =\eta^{\mu\nu}F_{a}+\frac{\tilde{k}^{\mu}\tilde{k}^{\nu}}{\tilde{k}^{2}}G_{a},\label{eq:C.2}
\end{align}
where we have introduced the following quantities,
\begin{align}
\left\{ F_{a},G_{a}\right\}  & =\frac{i\left(-\right)^{a-1}}{\left(4\pi\right)^{\frac{\omega}{2}}}\frac{1}{\Gamma\left(a\right)}\frac{1}{\left(s^{2}\right)^{a-1-\frac{\omega}{2}}}~\left\{ f_{a},g_{a}\right\} ,\label{eq:C.3}
\end{align}
with
\begin{align}
f_{a} & =\left(\frac{s\left|\tilde{k}\right|}{2}\right)^{a-1-\frac{\omega}{2}}K_{a-1-\frac{\omega}{2}}\left(\left|\tilde{k}\right|s\right),\label{eq:C.3a}\\
g_{a} & =\left(2a-2-\omega\right)\left(\frac{s\left|\tilde{k}\right|}{2}\right)^{a-1-\frac{\omega}{2}}K_{a-1-\frac{\omega}{2}}\left(\left|\tilde{k}\right|s\right)-2\left(\frac{s\left|\tilde{k}\right|}{2}\right)^{a-\frac{\omega}{2}}K_{a-\frac{\omega}{2}}\left(\left|\tilde{k}\right|s\right).\label{eq:C.3b}
\end{align}
Finally, we have
\begin{align}
\int\frac{d^{\omega}q}{\left(2\pi\right)^{\omega}}\frac{q^{\mu}q^{\nu}q^{\lambda}q^{\beta}}{\left(q^{2}-s^{2}\right)^{a}}e^{ik\wedge q} & =\left(\eta^{\mu\nu}\eta^{\lambda\beta}+\eta^{\nu\lambda}\eta^{\mu\beta}+\eta^{\nu\beta}\eta^{\lambda\mu}\right)H_{a}+\frac{\tilde{k}^{\lambda}\tilde{k}^{\beta}\tilde{k}^{\nu}\tilde{k}^{\mu}}{\tilde{k}^{4}}J_{a}\nonumber \\
 & +\left(\eta^{\lambda\beta}\frac{\tilde{k}^{\nu}\tilde{k}^{\mu}}{\tilde{k}^{2}}+\eta^{\nu\lambda}\frac{\tilde{k}^{\beta}\tilde{k}^{\mu}}{\tilde{k}^{2}}+\eta^{\nu\beta}\frac{\tilde{k}^{\lambda}\tilde{k}^{\mu}}{\tilde{k}^{2}}+\eta^{\lambda\mu}\frac{\tilde{k}^{\beta}\tilde{k}^{\nu}}{\tilde{k}^{2}}+\eta^{\beta\mu}\frac{\tilde{k}^{\lambda}\tilde{k}^{\nu}}{\tilde{k}^{2}}+\eta^{\nu\mu}\frac{\tilde{k}^{\lambda}\tilde{k}^{\beta}}{\tilde{k}^{2}}\right)I_{a},\label{eq:C.4}
\end{align}
where the quantities $H_{a}$, $I_{a}$ and $J_{a}$
are defined as the following,
\begin{align}
\left\{ H_{a},I_{a},J_{a}\right\}  & =-\frac{i\left(-\right)^{a}}{\left(4\pi\right)^{\frac{\omega}{2}}}\frac{1}{\Gamma\left(a\right)}\frac{1}{\left(s^{2}\right)^{a-2-\frac{\omega}{2}}}\frac{1}{s^{2}\tilde{k}^{2}}~\left\{ h_{a},i_{a},j_{a}\right\} ,\label{eq:C.5}
\end{align}
with
\begin{align}
h_{a} & =\left(2a-2-\omega\right)\left(\frac{s\left|\tilde{k}\right|}{2}\right)^{a-1-\frac{\omega}{2}}K_{a-1-\frac{\omega}{2}}\left(\left|\tilde{k}\right|s\right)-2\left(\frac{s\left|\tilde{k}\right|}{2}\right)^{a-\frac{\omega}{2}}K_{a-\frac{\omega}{2}}\left(\left|\tilde{k}\right|s\right),\label{eq:C.6a}\\
i_{a} & =\left[\left(2a-2-\omega\right)\left(2a-4-\omega\right)+s^{2}\tilde{k}^{2}\right]\left(\frac{s\left|\tilde{k}\right|}{2}\right)^{a-1-\frac{\omega}{2}}K_{a-1-\frac{\omega}{2}}\left(\left|\tilde{k}\right|s\right)\nonumber \\
 & -2\left(2a-4-\omega\right)\left(\frac{s\left|\tilde{k}\right|}{2}\right)^{a-\frac{\omega}{2}}K_{a-\frac{\omega}{2}}\left(\left|\tilde{k}\right|s\right),\label{eq:C.6b}\\
j_{a} & =\left(2a-4-\omega\right)\left[\left(2a-2-\omega\right)\left(2a-6-\omega\right)+2s^{2}\tilde{k}^{2}\right]\left(\frac{s\left|\tilde{k}\right|}{2}\right)^{a-1-\frac{\omega}{2}}K_{a-1-\frac{\omega}{2}}\left(\left|\tilde{k}\right|s\right)\nonumber \\
 & -2\left(2\left(2a-4-\omega\right)+s^{2}\tilde{k}^{2}\right)\left(\frac{s\left|\tilde{k}\right|}{2}\right)^{a-\frac{\omega}{2}}K_{a-\frac{\omega}{2}}\left(\left|\tilde{k}\right|s\right).\label{eq:C.6c}
\end{align}
%%%%%%%%%%%%%%%%%%%%%%%%%%%%%%%%%%%%%%%%%%%%%%%%%%%%%%%%%%%%%%%%%%%%%%%%%%%%%%%%%%%%%%%%%%%%%%%%%%%%%%%%%%%%%%%%%%%%%%%%%%%%%%%%%%%%%%%%%%%%%%%%%%%%%%%%%%%%%%%%%%%%%%%%%%%%%%%%%%%%%%%%%%%%%%%%%%%%%%%%%%%%%%%%%%%%%%
\section{One-loop form factors}
\label{appD}

In this section, we write down explicitly some lengthy expressions from the planar and nonplanar parts
from the self-energy form factors, discussed in Sec. \ref{sec:4}.
First, for the planar contribution of the transverse part $\left(\Pi_{\mathrm{e}}\right)$, Eq.\eqref{eq:3.3}, we immediately find
\begin{align}
\left(\Pi_{\mathrm{e}}\right)_{\mathrm{p}}\left(p^{2}\right) & =-\frac{e^{2}}{16\pi}\biggl\{-\frac{1}{ p^2}\int d\Phi\frac{1}{\sqrt{\Delta_1 ^2}}
\left(6p^{2}+16m^{2}-4\Delta_{1}^{2}\right)\nonumber \\
 & -\frac{1}{2 p^2}\int d\varUpsilon \frac{1}{\left(\Delta_2 ^2\right)^{\frac{3}{2}}}
 \biggl(16\left(y+z\right)^{2}p^{4}-10\left(y+z\right)p^{4}+p^{4} \nonumber \\
& +\left(26\left(y+z\right)-12\right)m^{2}p^{2}-16m^{4}-2\left[\left(7-2\left(y+z\right)\right)p^{2}+8m^{2}\right]\Delta_{2}^{2}\biggr)\nonumber \\
 & +\frac{1}{4}\int d\varXi\frac{1}{\left(\Delta_3 ^2\right)^{\frac{5}{2}}} \biggl(3p^{2}\left[4\left(z+w\right)\left(1-2\left(z+w\right)\right)m^{2}+\left(z+w\right)^{2}\left(3-4\left(z+w\right)\right)p^{2}\right]\nonumber \\
 & +\left[14\left(z+w\right)m^{2}+\left(4\left(z+w\right)+4\left(z+w\right)^{2}-1\right)p^{2}\right]\Delta_{3}^{2}-4\Delta_{3}^{4}\biggr)\biggr\},\label{eq:3.5}
\end{align}
while for the nonplanar contribution of \eqref{eq:3.3}, we obtain
\begin{align}
\left(\Pi_{\mathrm{e}}\right)_{\mathrm{n-p}}\left(p^{2}\right) & =\frac{e^{2}}{16\pi }\biggl\{-\frac{1}{ p^2}\int d\Phi
\frac{e^{-\Delta_{1}\left|\tilde{p}\right|}}{\sqrt{\Delta_1 ^2}} \left[\left(6p^{2}+16m^{2}\right)-4\Delta_{1}^{2}\right]\label{eq:3.6}\\
 & -\frac{1}{2 p^2}\int d\varUpsilon \frac{e^{-\Delta_{2}\left|\tilde{p}\right|}}{\left(\Delta_2 ^2\right)^{\frac{3}{2}}}\biggl(-\left[16p^{2}+\left(\left(4\left(y+z\right)+2\right)p^{2}-16m^{2}\right)\left[-1+\left|\tilde{p}\right|\Delta_{2}\right]\right]\Delta_{2}^{2} \nonumber \\
 &+\left[16\left(y+z\right)^{2}p^{4}-10\left(y+z\right)p^{4}+\left(26\left(y+z\right)-12\right)m^{2}p^{2}+p^{4}-16m^{4}\right]\left(1+\Delta_{2}\left|\tilde{p}\right|\right)  \biggr)\nonumber \\
 & + \frac{1}{4}\int d\varXi \frac{e^{-\Delta_{3}\left|\tilde{p}\right|}}{\left(\Delta_3 ^2\right)^{\frac{5}{2}}}\biggl(-\left[3p^{2}-8m^{2}-12\left(z+w\right)p^{2}\right]\left(1+\Delta_{3}\left|\tilde{p}\right|\right)\Delta_{3}^{2}+4\Delta_{3}^{4}\left(\Delta_{3}\left|\tilde{p}\right|-1\right)\nonumber \\
 &+\left[4\left(z+w\right)\left(1-2\left(z+w\right)\right)m^{2}+\left(z+w\right)^{2}\left(3-4\left(z+w\right)\right)p^{2}\right]\left(3+3\Delta_{3}\left|\tilde{p}\right|+\Delta_{3}^{2}\tilde{p}^{2}\right)p^{2}\nonumber \\
 & -\left(\left(8-14\left(z+w\right)\right)m^{2}-4\left(z+w\right)^{2}p^{2}-2p^{2}+8\left(z+w\right)p^{2}\right)\left[1+\Delta_{3}\left|\tilde{p}\right|-\tilde{p}^{2}\Delta_{3}^{2}\right]\Delta_{3}^{2} \biggr)\biggr\}.\nonumber
\end{align}

Moreover, similar expressions follow for the NC transverse part $\left(\widetilde{\Pi}_{\mathrm{e}}\right)$, Eq.\eqref{eq:3.7}.
Without any complication, the planar contribution results in
\begin{align}
\left(\widetilde{\Pi}_{\mathrm{e}}\right)_{\mathrm{p}}\left(p^{2}\right) & =-\frac{e^{2}}{16\pi }\biggl\{\frac{1}{p^{2}}\int d\Phi\frac{1}{\sqrt{\Delta_{1}^2}} \left(\left(16m^{2}-4p^{2}\right)-2\Delta_{1}^{2}\right)\label{eq:3.8} \\
 & -\frac{1}{2p^{2}}\int d\varUpsilon\frac{1}{\left(\Delta_{2}^{2}\right)^{\frac{3}{2}}}
 \biggl(m^{2}\left(-30\left(y+z\right)p^{2}+7p^{2}+16m^{2}\right)-3p^{4}+10\left(y+z\right)p^{4}\nonumber \\
 &-2\left(2p^{2}+4\left(y+z\right)p^{2}-16m^{2}\right)\Delta_{2}^{2}\biggr) \nonumber \\
 & +\frac{1}{4} \int d\varXi \frac{1}{\left(\Delta_{3}^{2}\right)^{\frac{5}{2}}}\biggl(3p^{2}\left[-4m^{2}\left(z+w\right)+7m^{2}\left(z+w\right)^{2}+\left(z+w\right)^{2}p^{2}-4\left(z+w\right)^{3}p^{2}\right]\nonumber \\
 & -\left[8\left(z+w\right)^{2}p^{2}+5p^{2}-28\left(z+w\right)p^{2}-m^{2}\bigg[9-28\left(z+w\right)\bigg]\right]\Delta_{3}^{2}+8\Delta_{3}^{4}\biggr)\biggr\},\nonumber
\end{align}
and the nonplanar contribution reads
\begin{align}
\left(\widetilde{\Pi}_{\mathrm{e}}\right)_{\mathrm{n-p}}\left(p^{2}\right) & =\frac{e^{2}}{16\pi}\biggl\{\frac{1}{p^{2}}\int d\Phi \frac{e^{-\Delta_{1}\left|\tilde{p}\right|}}{\sqrt{\Delta_{1}^2}} \left[\left(-4p^{2}+16m^{2}\right)-4\frac{1}{\left|\tilde{p}\right|}\Delta_{1}-6\Delta_{1}^{2}\right]\label{eq:3.9} \\
 & -\frac{1}{2 p^{2}}\int d\varUpsilon \frac{e^{-\Delta_{2}\left|\tilde{p}\right|}}{\left(\Delta_{2}^{2}\right)^{\frac{3}{2}}} \biggl(-\left(4p^{2}+8\left(y+z\right)p^{2}-32m^{2}\right)\left[1-\Delta_{2}\left|\tilde{p}\right|\right]\Delta_{2}^{2} \nonumber \\
 &+\left[m^{2}\left(-30\left(y+z\right)p^{2}+7p^{2}+16m^{2}\right)-3p^{4}+10\left(y+z\right)p^{4}\right]\left(1+\Delta_{2}\left|\tilde{p}\right|\right)\biggr)\nonumber \\
 & +\frac{1}{4}\int d\varXi \frac{e^{-\Delta_{3}\left|\tilde{p}\right|}}{\left(\Delta_{3}^{2}\right)^{\frac{5}{2}}} \biggl(-\left(7m^{2}+p^{2}-12\left(z+w\right)p^{2}\right)\left(1+\Delta_{3}\left|\tilde{p}\right|\right)\Delta_{3}^{2}+8\Delta_{3}^{4}\left(1-\Delta_{3}\left|\tilde{p}\right|\right)\nonumber \\
&+ p^{2}\left[-4m^{2}\left(z+w\right)+7m^{2}\left(z+w\right)^{2}+\left(z+w\right)^{2}p^{2}-4\left(z+w\right)^{3}p^{2}\right]\left(3+3\Delta_{3}\left|\tilde{p}\right|+\Delta_{3}^{2}\tilde{p}^{2}\right)\nonumber \\
 & -\left[8\left(z+w\right)^{2}p^{2}+4p^{2}-16\left(z+w\right)p^{2}-2m^{2}\bigg[8-14\left(z+w\right)\bigg]\right]\left(1+\Delta_{3}\left|\tilde{p}\right|-\tilde{p}^{2}\Delta_{3}^{2}\right)\Delta_{3}^{2}\biggr)\biggr\}.\nonumber
\end{align}

Finally, for the CP odd part $\left(\Pi_{\mathrm{o}}^{\mathrm{\tiny{A}}}\right)$, Eq.\eqref{eq:3.10}, we find the planar part of the expression
\begin{align}
\left(\Pi_{\mathrm{o}}^{\mathrm{\tiny{A}}}\right)_{\mathrm{p}}\left(p^{2}\right) & =-\frac{me^{2}}{8\pi}\biggl\{\int d\Phi\frac{5}{\sqrt{\Delta_{1}^2}}-\frac{1}{2}\int d\varUpsilon\frac{1}{\left(\Delta_{2}^2\right)^{\frac{3}{2}}}
\left(\left(\left(-6+15x-5x^{2}\right)p^{2}+7m^{2}\right)+5\Delta_{2}^{2}\right)\nonumber \\
 & -\frac{1}{4}\int d\varXi\frac{1}{\left(\Delta_{3}^2\right)^{\frac{5}{2}}}\biggl(3\left(z+w\right)^{2}\left\{ 2m^{2}-5\left(z+w\right)p^{2}+4p^{2}\right\} p^{2}\nonumber \\
 &-\left(2m^{2}+4p^{2}-15\left(z+w\right)p^{2}\right)\Delta_{3}^{2}\biggr)\biggr\},\label{eq:3.11}
\end{align}
whereas, for the nonplanar part, we obtain the following contribution:
\begin{align}
\left(\Pi_{\mathrm{o}}^{\mathrm{\tiny{A}}}\right)_{\mathrm{n-p}}\left(p^{2}\right) & =\frac{me^{2}}{8\pi}\biggl\{5\int d\Phi\frac{e^{-\Delta_{1}\left|\tilde{p}\right|}}{\sqrt{\Delta_{1}^2}}+\frac{1}{2}\int d\varUpsilon \frac{e^{-\Delta_{2}\left|\tilde{p}\right|}}{\left(\Delta_{2}^2\right)^{\frac{3}{2}}}
\left(\left(\left(6-15x+5x^{2}\right)p^{2}-7m^{2}\right)\left(1+\Delta_{2}\left|\tilde{p}\right|\right)-5\Delta_{2}^{2}\right)\nonumber \\
 & -\frac{1}{4}\int d\varXi \frac{e^{-\Delta_{3}\left|\tilde{p}\right|}}{\left(\Delta_{3}^2\right)^{\frac{5}{2}}} \biggl(\left(z+w\right)^{2}\left\{ 2m^{2}+\left(4-5z-5w\right)p^{2}\right\} \left(3+3\Delta_{3}\left|\tilde{p}\right|+\Delta_{3}^{2}\tilde{p}^{2}\right)p^{2}\nonumber \\
 &-\left(2m^{2}+4p^2-15\left(z+w\right)p^{2}\right)\left(1+\Delta_{3}\left|\tilde{p}\right|\right)\Delta_{3}^{2}\biggr)\biggr\}.\label{eq:3.12}
\end{align}

%%%%%%%%%%%%%%%%%%%%%%%%%%%%%%%%%%%%%%%%%%%%%%%%%%%%%%%%%%%%%%%%%%%%%%%%%%%%%%%%%%%%%%%%%%%%%%%%%%%%%%%%%%%%%%%%%%%%%%%%%%%%%%%%%%%%%%%%%%%%%%%%%%%%%%%%%%%%%%%%%%%%%%%%%%%%%%%%%%%%%%%%%%%%%%%%%%%%%%%%%%%%%%%%%%%%%%

\global\long\def\link#1#2{\href{http://eudml.org/#1}{#2}}
 \global\long\def\doi#1#2{\href{http://dx.doi.org/#1}{#2}}
 \global\long\def\arXiv#1#2{\href{http://arxiv.org/abs/#1}{arXiv:#1 [#2]}}
 \global\long\def\arXivOld#1{\href{http://arxiv.org/abs/#1}{arXiv:#1}}

%%%%%%%%%%%%%%%%%%%%%%%%%%%%%%%%%%%%%%%%%%%%%%%%%%%%%%%%%%%%%%%%%%%%%%%%%%%%%%%%%%%%%%%%%%%%%%%%%%%%%%%%%%%

\end{document}